\documentclass[a4paper,11pt]{article}
\usepackage{jcappub,tablefootnote,adjustbox}  
\usepackage{lineno}
\usepackage{cancel,ulem}\usepackage{xcolor}
\newcommand{\bea}{\begin{eqnarray}}
 
\newcommand{\eea}{\end{eqnarray}}
\newcommand{\bma}{\begin{pmatrix}}
\newcommand{\ema}{\end{pmatrix}}
\newcommand{\be}{\begin{equation}}
\newcommand{\ee}{\end{equation}}
\newcommand{\beno}{\begin{equation*}}
\newcommand{\eeno}{\end{equation*}}

\def\doi{http://doi.org}

\arxivnumber{2412.05382} % Only if you have one
\title{Scalar perturbation and density contrast evolution in $f(Q,C)$ gravity}

\author[a]{Ganesh Subramaniam,}
\author[b]{Avik De,}
\author[b]{Tee-How Loo}
\author[a]{and Yong Kheng Goh}
\affiliation[a]{Department of Mathematical and Actuarial Sciences\\ Universiti Tunku Abdul Rahman, Jalan Sungai Long,
43000 Cheras, Malaysia}
\affiliation[b]{Institute of Mathematical Sciences, Faculty of Science, Universiti Malaya, 50603 Kuala Lumpur, Malaysia}

\emailAdd{ganesh03@1utar.my, avikde@um.edu.my, looth@um.edu.my, gohyk@utar.edu.my}

\note{The research was supported by the Ministry of Higher Education (MoHE), through the Fundamental Research Grant Scheme (FRGS/1/2023/STG07/UM/02/3, project no.: FP074-2023).}

\abstract{
The symmetric teleparallel theory offers an alternative gravitational formulation which can elucidate events in the early and late universe without requiring the physical existence of dark matter or dark energy. In this formalism, $f(Q, C)$ gravity has been recently introduced by incorporating the boundary term $C$ with the non-metricity scalar $Q$. In this paper, we develop the theory of cosmological scalar perturbation for $f(Q, C)$ gravity, and retrieve that of $f(\mathring{R})$ and $f(Q)$ gravity from our result. The analysis assumes a model-independent approach within these theories that adheres to the conventional continuity equation at the background level. We derive the density contrast equation by employing some standard cosmological approximations, where the $f(Q,C)$ theory is encoded in the effective Newtonian constant $G_{eff}$. Finally, we derive the evolution equation of density growth $f_g$. }
\begin{document}
\maketitle
\flushbottom
\section{Introduction}
Modern cosmological research is mostly based on the $\Lambda$CDM model in which General Relativity (GR) governed by the Einstein's field equation (EFE) is modified by including a cosmological constant $\Lambda$ to explain the currently observed accelerated expansion of the universe. Unfortunately, the $\Lambda$CDM model is not free from criticism. The most significant ones are the cosmological constant problem - the observed value of vacuum energy being 120 orders of magnitude different from the one computed theoretically \cite{adler1995}, the coincidence problem - the observed density of dark energy and dark matter having the same order of magnitude at the beginning of the present epoch at redshift $z\approx 0.55$ \cite{velten2014} and the Hubble tension problem - the different methods of measurement yielding different results for the Hubble parameter, $H$ at present \cite{perivolaropoulos2022}. These fundamental issues warrant an improvement of the $\Lambda$CDM model. There are basically two ways one can modify GR, first by introducing some additional matter components as in the case of $\Lambda$CDM model, or by revising the geometrical formulation of the theory. In this paper, we investigate a modified gravity theory of second type, the so-called $f(Q, C)$ gravity under symmetric teleparallelism, in which the non-metricity character of the spacetime is associated with the gravitational interaction instead of the traditional curvature based theories.
\par
The $f(\mathring{R})$ theory is a simple modification to GR by replacing the Ricci scalar $\mathring{R}$ by arbitrary function $f(\mathring{R})$ in the gravitational action of GR \cite{buchdahl1970}. It is a quantum mechanics compatible, higher-order derivative theory developed to solve the renormalization problems of GR \cite{sotiriou2010}. De Felice and Tsujikawa \cite{felice2010} reviewed various aspects of $f(\mathring R)$ theories in detail; dark energy, local gravity constraints and cosmological perturbation being some particularly interesting ones. Several other curvature-based modifications of GR were proposed in the literature, like $f(\mathring{R}, G)$ gravity \cite{nojiri2005}, $f(\mathring{R}, L_m)$ gravity \cite{harko2010}, $f(\mathring{R}, \mathbb T)$ gravity \cite{harko2011}, among others.
% \par

\par
Historically, cosmological perturbation theory has been introduced to bridge together the early fluctuation in the energy density of the universe to the formation of the structure such as formation of clusters of galaxy \cite{kodama1984, mukhanov1992, malik2009}. It is known from the observed CMB at redshift $z\approx 1000$, the temperature fluctuation is about $\delta \text{T/T}\approx 10^{-5}$ which is of the same order with the fluctuation of baryonic density $\delta\rho/\bar{\rho}$ \cite{plank2018}. From this time ($z\approx 1000$) onward, the growth of density perturbation up to the linear regime is described by well-known density contrast evolution equation \cite{peebles1980, gleyzes2016} 
\begin{equation}
\ddot{\delta}(t)+2H(t)\dot{\delta}(t)=4\pi G\bar\rho\delta(t)\,,\label{eeq}
\end{equation}
where $H(t)$ is Hubble parameter, $\bar\rho$ is the background energy density and $G$ is the Newtonian constant. Though an initial perturbation of order $10^{-5}$ is sufficient to describe the formation of structures as observed today but the growth of the factor of $10^3$ since the time of recombination leads to the perturbation of order $10^{-2}$ which is smaller than the nonlinear structures observed at the present \cite{ostriker1993}, one of the reasons that motivates researcher to explore perturbation theory in the modified gravity theories. In modified theories of gravity, the Newtonian constant $G$ in equation (\ref{eeq}) is replaced with the effective gravitational constant $G_{eff}$ to indicate the modifications in the respective modified gravity theories \cite{tsujikawa2007, tsujikawa2008,dombriz2008, zheng2011, mirzatuny2014, bahamonde2021}.
\par
Of late, researchers attempted to consider the geometric trinity- Curvature, Torsion and Non-metricity characterisitics of a spacetime into account while investigating gravitational interaction. Metric and symmetric teleparallel theories of gravity were introduced, and in the respective arena, the modified $f(T)$ ($T$ the torsion) and $f(Q)$ ($Q$ the non-metricity scalar) theories of gravity (for a brief review see \cite{cai2016,bahamonde2023,heisenberg2024}). Originally, these theories were introduced to understand recent accelerated cosmic expansion \cite{myrzakulov2011,solanki2021,koussour2022,narawade2023,goswami2023,subramaniam2023,subramaniam2023a,avik2023aa}, to explore Hubble tensions \cite{nunes2018} etc, and most part of the literature is focused on the evolution of the background spacetime. At background level, these two theories produce the same set of equations of motion when vanishing affine connection is used in the formulation. However, the perturbation equations are significantly different. Furthermore, a specific model of the $f(Q)$ gravity e.g. $f(Q)=Q+\beta\sqrt{-Q}$ is indistinguishable from $\Lambda$CDM model at the background spacetime, but notably different in the perturbative level \cite{frusciante2021}. The matter power spectrum and lensing effect on CMB angular power spectrum were suppressed and further enhanced integrated Sach-Wolfe tail of CMB temperature anisotropies. For further reading on cosmological perturbation in the modified gravity theories, please see \cite{chen2011,zheng2011,jimenez2020,atayde2021,anagnostopoulos2021,frusciante2021,albuquerque2022} and the references therein. \par
It is worth pointing out that $f(Q)$ gravity bears some type of ghost instability in the sense that the scalar mode appears with a negative kinetic energy \cite{hu2022, tomonari2024}. Nevertheless, the `teleparallel theories of gravity' are not entirely doomed by this discovery. First of all, this is not fitting to rely solely on linear perturbation. One must calculate higher-order perturbations prior to drawing a definitive conclusion, which is, of course, exceedingly challenging \cite{saha2025}. Very recently, works were published to contact ghost-free models. For example, the connections incorporated by different couplings of matter field may survive from such inconsistencies \cite{ambrosio2023,heisenberg2024, heisenberg2024b, bello2024}. Subsequently, it is demonstrated in metric teleparallel counterpart that the issue of the ghost may be mitigated by incorporating a non-minimally connected scalar field, even at the linear perturbation level \cite{hu2024}. Besides, it has been shown that the ghost scalar field resided in $f(Q)$ gravity can be removed by some constraint, since, it does not propagate, thus the $f(Q)$ theory is healthy and free from such ghost \cite{hu2023}.\\

\par
 In the present work, we examine a category of modified gravity theories whereby the gravitational action incorporates a comprehensive function $f(Q, C)$, with $Q$ and $C$ representing the non-metricity scalar and its deviation from the conventional Levi-Civita based Ricci scalar $\mathring{R}$, respectively. This kind of modified symmetric teleparallel gravity was initially presented in \cite{capozzielo2023,avik2023a}, along with its basic cosmological formulations, which were explored in three distinct affine connection situations. Subsequently, the $\Lambda$CDM reconstruction of such theories has been carried out in \cite{gadbail2023}. Several cosmological aspects of $f(Q,C)$ theory have been studied, in both theoretical and observational grounds, for reference see \cite{samaddar2025, sadatian2024, alruwaili2025, bhoyar2025, myrzakulov2025, chandra2024, zotos2024, mmaurya2024, maurya2024, de2023, paliathanasiss2024, dixit2024, usman2024, murtaza2025}. In this theory, the boundary term $C$ represents the fourth-order contributions to the field equations, like the $f(\mathring{R})$ theories of gravity and unlike the second order field equations of GR. 

This paper is organized as follows:
\par
We begin with a brief introduction of $f(Q, C)$ gravity in Section \ref{II}, followed by a description of the background evolution in FLRW-spacetime in Section \ref{III}. In Section \ref{IV}, we develop the scalar metric perturbation for $f(Q, C)$ gravity. Next, in Section \ref{V}, we discuss the perturbation of the connection field equation and conservation equation and derive the expression of $G_{eff}$ for $f(Q, C)$ gravity. In Section \ref{VI}, we discuss density growth and growth index parameters along with figures comparing a few $f(Q,C)$ models with $\Lambda$CDM. Finally, in section \ref{VII} we conclude the paper. The detailed calculations are provided in separate appendices at the end. 

\par
In our notation, the Greek indices run from $0$ to $3$ representing spacetime coordinates, while Latin indices run from $1$ to $3$ representing spatial coordinates. We also used the convention $\kappa^2=8\pi G$. The quantities $\mathring{( \quad )}$ are calculated with respect to the Levi-Civita connection, $\mathring{\Gamma}$.
%%%%%%%%%%%%%%%%%%%%%%%%%%%%%%%%%%%%%%%%%%%%%%%%%%%%%%%%%%%%%%%%%%%%%%%%%%%%
\section{Fundamentals of the $f(Q, C)$ gravity}\label{II}
%%%%%%%%%%%%%%%%%%%%%%%%%%%%%%%%%%%%%%%%%%%%%%%%%%%%%%%%%%%%%%%%%%%%%%%%%%%%
 In this section, we introduce the notation of the variables, summarise the $f(Q)$ theory and introduce the boundary term $C$ to describe the $f(Q, C)$ theory as in \cite{avik2023a, capozzielo2023}. Suppose $\mathcal{M}$ is a 4-dimensional Lorentzian manifold with coordinates $(t, x^1,x^2, x^3)$ and endowed with a metric tensor $g_{\mu\nu}$. In addition we consider a general affine connection $\Gamma^\alpha{}_{\mu\nu}$, 
 such that the non-metricity tensor
 \begin{align}
 Q_{\lambda\mu\nu}\equiv\nabla_\lambda g_{\mu\nu}=\partial_\lambda g_{\mu\nu}-\Gamma^\beta{}_{\lambda\mu} g_{\beta\nu}-\Gamma^\beta{}_{\lambda\nu}g_{\mu\beta}\,,\label{cd}
 \end{align}
 is non-zero on the manifold $\mathcal{M}$. In symmetric teleparallel gravity, we have vanishing torsion (due to $\Gamma^\alpha{}_{\mu\nu}=\Gamma^\alpha{}_{\nu\mu}$) and curvature ($R^\lambda{}_{\mu\alpha\nu}=0$) postulates.
\par
The deviation of this general connection $\Gamma^\alpha{}_{\mu\nu}$ from the symmetric and metric-compatible Levi-Civita connection $\mathring{\Gamma}^\alpha{}_{\mu\nu}=\frac{1}{2}g^{\alpha\beta}(\partial_\nu g_{\beta\mu}+\partial_\mu g_{\beta\nu}-\partial_\beta g_{\mu\nu})$   %\,,\label{ac}
can be written as
 \begin{align}
\Gamma^\alpha{}_{\mu\nu}=\mathring{\Gamma}^\alpha{}_{\mu\nu}+L^\alpha{}_{\mu\nu}\,,\label{ac1}
 \end{align}
 where $L^\alpha{}_{\mu\nu}$ is the disformation tensor given by
 \begin{align}
     L^\lambda{}_{\mu\nu}=\frac{1}{2}(Q^\lambda{}_{\mu\nu}-Q_\mu{}^\lambda{}_\nu-Q_\nu{}^\lambda{}_\mu)\,.\label{L}
 \end{align}
 We can construct two non-metricity vectors
 \begin{align}
     Q_\mu=g^{\nu\lambda}Q_{\mu\nu\lambda}=Q_\mu{}^\nu{}_\nu\,,\qquad\qquad \tilde{Q}_\mu=g^{\nu\lambda}Q_{\nu\mu\lambda}=Q_{\nu\mu}{}^\nu\,,\label{vq}
 \end{align}
 and similarly
 \begin{align}
     L_\mu=L_\mu{}^\nu{}_\nu\,,\qquad\qquad \tilde{L}_\mu=L_{\nu\mu}{}^\nu\,.\label{vl}
 \end{align}
The superpotential tensor is given by
 \begin{align}
     P^\lambda{}_{\mu\nu}=\frac{1}{4}\left(-2L^\lambda{}_{\mu\nu}+(Q^\lambda-\tilde{Q}^\lambda) g_{\mu\nu}-\delta^\lambda{}_{(\mu}Q_{\nu)}\right)\,,
 \end{align}
and the non-metricity scalar as
\begin{align}
    Q=Q_{\alpha\beta\gamma}P^{\alpha\beta\gamma}\,.\label{q}
\end{align}
We can obtain the following relations with the vanishing torsion and curvature constraints
\begin{align}
 \mathring{R}_{\mu\nu}+\mathring{\nabla}_\alpha L^\alpha{}_{\mu\nu}-\mathring{\nabla}_\nu \tilde{L}_\mu+\tilde{L}_\alpha L^\alpha{}_{\mu\nu}-L_{\alpha\beta\nu}L^{\beta\alpha}{}_\mu&=0\,,\label{ce1}\\
 \mathring{R}+\mathring{\nabla}_\alpha(L^\alpha-\tilde{L}^\alpha)-Q&=0\,.\label{ce2}
\end{align}
The equation (\ref{ce2}) can be used as the definition of the boundary term $C$
\begin{align}
    C\equiv \mathring{R}-Q=-\mathring{\nabla}_\alpha(L^\alpha-\tilde{L}^\alpha)=-\mathring{\nabla}_\alpha(Q^\alpha-\tilde{Q}^\alpha)\,.\label{c}
\end{align}

The gravitational action for the $f(Q, C)$ gravity can be expressed as  
\begin{align}
    S = \frac1{2\kappa^2}\int f(Q,C) \sqrt{-g}\,d^4 x+\int \mathcal{L}_m \sqrt{-g}\,d^4 x\,,\label{A}
\end{align}
analogous to the $f(\mathring{R})$ gravity and the $f(Q)$ gravity where $g$ is the metric determinant, $\mathcal{L}_m$ is matter Lagrangian and $\kappa^2=8\pi G$. 
The variation of action (\ref{A}) with respect to the metric tensor produces the corresponding field equations \cite{avik2023a},
\begin{align} \label{FE1}
   f_Q \mathring{G}_{\mu\nu}+\frac{1}{2} g_{\mu\nu} (Qf_Q-f) &+ 2f_{QQ} P^\lambda{}_{\mu\nu} \mathring{\nabla}_\lambda Q -2P^\lambda{}_{\mu\nu}\nabla_\lambda f_C\notag\\&+\left(\frac{C}{2}g_{\mu\nu}-\mathring{\nabla}_\mu\mathring{\nabla}_\nu+g_{\mu\nu}\mathring{\nabla}^\alpha\mathring{\nabla}_\alpha\right)f_C= \kappa^2 \mathcal{T}_{\mu\nu}\,,
\end{align}
where $f_Q =\frac{\partial f}{\partial Q}$, $f_{QQ}=\frac{\partial^2 f}{\partial Q^2}$, $f_C =\frac{\partial f}{\partial C}$, $f_{CC}=\frac{\partial^2 f}{\partial C^2}$.
The energy-momentum tensor for a perfect fluid is $\mathcal{T}_{\mu\nu}=(p+\rho)u_\mu u_\nu+pg_{\mu\nu}$, where $p, \rho$ and $u_\mu$ are pressure, energy density and the fluid 4-velocity respectively. 

It is important to keep in mind that, unlike GR, in the teleparallel theory the affine connection is independent of the metric tensor, and so the teleparallel theories are technically metric-affine theories where both the metric and the connection act as dynamic variables. Therefore, by taking variation of the action (\ref{A}) with respect to the affine connection, 
we obtain the connection field equation
\begin{align}\label{eqn:FE2-invar}
(\nabla_\mu-\tilde L_\mu)(\nabla_\nu-\tilde L_\nu)
\left[4(f_Q-f_C)P^{\mu\nu}{}_\lambda+\Delta_\lambda{}^{\mu\nu}\right]=0\,,
\end{align}
where $\Delta_\lambda{}^{\mu\nu}$ is the hypermomentum tensor defined as \cite{hehl1976}
$\Delta_\lambda{}^{\mu\nu}=-\frac2{\sqrt{-g}}\frac{\delta(\sqrt{-g}\mathcal L_M)}{\delta\Gamma^\lambda{}_{\mu\nu}}\,.$

%%%%%%%%%%%%%%%%%%%%%%%%%%%%%%%%%%%%%%%%%%%%%%%%%%%%%%%%%%%%%%%%%%%%%%%%%%%%
\section{Background spacetime for $f(Q,C)$ gravity}\label{III}
%%%%%%%%%%%%%%%%%%%%%%%%%%%%%%%%%%%%%%%%%%%%%%%%%%%%%%%%%%%%%%%%%%%%
For the background FLRW metric
\begin{equation}\label{metric}
     ds^2=-dt^2+a(t)^2\delta_{ij}dx^idx^j\,,
 \end{equation} 
  the energy-momentum tensor for a perfect fluid $\bar {\mathcal{T}}_{\mu\nu}=(\bar \rho, a^2\bar p, a^2\bar p, a^2\bar p)$ is considered, where an overbar ($\bar{ \quad }$) denoting quantities belong to background spacetime. In this paper, we consider a vanishing affine connection ($\Gamma^\lambda{}_{\mu\nu}=0$) throughout. The non-metricity scalar $\bar Q=-6H^2$ and the boundary term $\bar C=6(3H^2+\dot{H})$ can be computed from equations (\ref{q}) and (\ref{c}) respectively. The Friedmann-like equations are obtained from equation (\ref{FE1}) as \cite{avik2023a} 
 \begin{align}
    3H^2=&\kappa^2(\bar \rho+\bar \rho^{DE})\,,\label{fle1}\\
    -(2\dot{H}+3H^2)=&\kappa^2(\bar p+\bar p^{DE})\,,\label{fle2}
 \end{align}
 where
  \begin{align}
     \kappa^2\bar \rho^{DE}=&3H^2(1-2f_Q)-\frac{f}{2}+3(3H^2+\dot{H})f_C-3H\dot{f}_C\,,\label{rode}\\
     \kappa^2\bar p^{DE}=&-2\dot{H}(1-f_Q)-3H^2(1-2f_Q)+\frac{f}{2}+2H\dot{f}_Q-3(3H^2+\dot{H})f_C+\ddot{f}_C\,.\label{pde}
 \end{align}
Here $\bar \rho$ and $\bar p$ are perfect fluid energy density and pressure for the background spacetime respectively. Equations (\ref{fle1}) and (\ref{fle2}) describe the dynamics of the universe under the $f(Q, C)$ theory. The additional terms $\bar \rho^{DE}$ and $\bar p^{DE}$ are the dark energy density and pressure, corresponding to the modification of GR promise to offer a reasonable negative thrust to generate the accelerated expansion of the late universe. Conservation of energy-momentum tensors, with the assumptions of vanishing hypermomentum tensor provides the usual continuity equation 
\begin{equation}
    \dot{\bar \rho}=-3H(1+\omega)\bar \rho\,.\label{ec}
\end{equation}

Using equations (\ref{fle1}), (\ref{rode}) and the energy density in matter-era relation $\bar \rho^m=\bar \rho_0 a^{-3}$ (which is obtainable by integrating (\ref{ec}) for $\omega=0$), we have 
\begin{align}
    2H^2{f}_Q+\frac{f}{6}-(3H^2+\dot{H})f_C+H\dot{f}_C=H_0{}^2\Omega^{m0}a^{-3}\,,\label{beq1}
\end{align}
where $\Omega^{m0}=\frac{\kappa^2\bar{\rho}_0}{3H_0{}^2}$ is the present value of the matter density parameter, and $H_0$ is the present value of the Hubble parameter.

%%%%%%%%%%%%%%%%%%%%%%%%%%%%%%%%%%%%%%%%%%%%%%%%%%%%%%%%%%%%%%%%%%%%%%%%%%%
\section{Metric perturbation for $f(Q,C)$ gravity}\label{IV}
%%%%%%%%%%%%%%%%%%%%%%%%%%%%%%%%%%%%%%%%%%%%%%%%%%%%%%%%%%%%%%%%%%%%%%%%%%%%%%%%%%%%%%%%%%%%%%%%%%%%%%%%
 
In this section, we examine the linear cosmological perturbation theory within the context of the $f(Q, C)$ gravity. 

We consider first the general metric perturbation of the form\footnote{Note that the components are $g_{00}=-(1+2A), \ g_{0i}=g_{i0}=-aB_i, \ g_{ij}=a^2\left[(1+2D)\delta_{ij}+2E_{ij}\right]$.}
\begin{equation}\label{e1}
    ds^2=-(1+2A)dt^2-2a B_i dx^idt+a^2\left[(1+2D)\delta_{ij}+2E_{ij}\right]dx^idx^j\,,
\end{equation}
where $A=\phi$, $D=-\psi+\frac{1}{3}\nabla^2E$ while $\phi$ and $\psi$ are scalar functions\footnote{In many literature $\psi$ is called curvature perturbation ($\psi=-D+\frac{1}{3}\nabla^2E$).} of $(t,x^1,x^2,x^3)$.
% and $B_i, E_{ij}$ are curl and divergence less quantities. 
It is common to perform scalar-vector-tensor (SVT) decomposition for $B_i$ and $E_{ij}$ \cite{kurkisuonio2024} \footnote{We followed \cite{kurkisuonio2024} notations, so our results might differ by a negative sign from some published works.}. First,

\begin{align}
    B_i=&B^S{}_i+\vec{B}_i=-\partial_iB+\vec{B}_i\,, \ \ \quad \nabla\times(B^S{}_i)=0\,, \ \ \ \ \nabla\cdot \vec{B}_i=0\,,\label{b}
\end{align}
where $B$ is a scalar function of $(t, x^1, x^2, x^3)$. Similarly,
\begin{align}
    E_{ij}=&E^S{}_{ij}+\vec{E}_{ij}+\vec{E}^T{}_{ij}\,,\label{e}
\end{align}
where $E^S{}_{ij}$ is the scalar term, $\vec{E}_{ij}$ the vector term and $\vec{E}^T{}_{ij}$ the tensor term which can be expressed as
\begin{align}
     E^S{}_{ij}=&\left(\partial_{ij}-\frac{1}{3}\delta_{ij}\nabla^2\right)E\,,\label{Es}\\
     \vec{E}_{ij}=&-\frac{1}{2}(\partial_i E_j+\partial_j E_i)\,, \ \ \ \ \delta^{ij}\partial_jE_i=\nabla\cdot\vec{E}=0\,,\\
     \delta^{ij}\partial_k E^T{}_{ij}=&0\,,\quad \delta^{ij}E^T{}_{ij}=0\,,
\end{align}
where $E$ is a scalar function of $(t, x^1, x^2, x^3)$.
In the following, we consider scalar metric perturbation. Hence, the vector and tensor contributions from $B_i$ and $E_{ij}$ are ignored. Therefore, by using $A=\phi$, $D=-\psi+\frac{1}{3}\nabla^2 E$, and the $B^S{}_i$ , $E^S{}_{ij}$ terms of equations (\ref{b}) and (\ref{e}), the metric (\ref{e1}) can be written as
\begin{equation}
    ds^2=-(1+2\phi)dt^2+2a \partial_iB dx^idt+a^2\left[(1-2\psi)\delta_{ij}+2\partial_{ij}E\right]dx^idx^j\,.\label{smp}
\end{equation}
For $f(Q, C)$ gravity, the first-order perturbation of the field equations (\ref{FE1}) can be written as
\begin{align}
    f_Q \delta \mathring{G}_{\mu\nu}+\mathring{\bar{G}}_{\mu\nu}&\delta f_Q+\frac{1}{2} (\bar Q f_Q-f)\delta g_{\mu\nu}+\frac{1}{2}\bar g_{\mu\nu}\delta (Q f_Q-f) + 2\delta(f_{QQ} P^\lambda{}_{\mu\nu} \mathring{\nabla}_\lambda Q)\notag\\&-2\delta\left[P^\lambda{}_{\mu\nu}\nabla_\lambda f_C\right]+\delta\left[\left(\frac{C}{2}g_{\mu\nu}-\mathring{\nabla}_\mu\mathring{\nabla}_\nu+g_{\mu\nu}\mathring{\nabla}^\alpha\mathring{\nabla}_\alpha\right)f_C\right] =\kappa^2 \delta \mathcal{T}_{\mu\nu}\,,\label{fefqc}
\end{align}
where the energy-momentum tensor perturbations for a perfect fluid is 
\begin{equation}
    \delta \mathcal{T}_{\mu\nu}=(\delta\rho+\delta p)\bar u_\mu\bar u_\nu+(\bar \rho+\bar p)(\bar u_\mu\delta u_\nu+\bar u_\nu\delta u_\nu)+\bar p\delta g_{\mu\nu}+\bar g_{\mu\nu}\delta p\,.\label{deltap}
\end{equation}
Here $\bar u_\mu$ and $\delta u_\mu$ are background fluid 4-velocity and its perturbation respectively.
Note that 4-velocity perturbation can be decomposed into
\begin{align}
    \delta u_\mu=-\partial_\mu u+\vec{u}_\mu\,.
\end{align}
Hence, for the scalar metric perturbation, the components of energy-momentum tensor perturbation (\ref{deltap}) are
\begin{align}\label{pemt1}
    \delta \mathcal{T}_{00}=&\delta\rho+2\bar{\rho}\phi\,, 
    % \quad\delta T_{0i}=-(\bar{\rho}+\bar{p})u_i\,,
    \quad\quad\quad\quad\quad\quad\quad\quad\quad\quad
    % \delta T_{0i}=\delta T_{i0}=&-(\bar{\rho}+\bar{p})u_i-a\bar p B_i,
    \delta \mathcal{T}_{0i}=\delta \mathcal{T}_{i0}=(\bar{\rho}+\bar{p})\partial_iu+a\bar p \partial_iB\,,
    \notag\\
    \sum_i\delta \mathcal{T}_{ii}=&3\left(2a^2\bar{p}\left(-\psi+\frac{1}{3}\nabla^2E\right)+a^2\delta p\right)\,,\quad\quad\delta \mathcal{T}_{ij}=2a^2\bar p\partial_{ij}E\,.
\end{align}
 It is obvious from the last term on the left-hand side (LHS) of (\ref{fefqc}), it is a set of fourth-order differential equations. From (\ref{fefqc}) and (\ref{pemt1}), we obtain the following perturbation quantities. Detailed calculations are provided in Appendix \ref{A00}.
\begin{align}
    \kappa^2 \delta\rho=&\left[-6H(\dot{\psi}+H\phi)+2\frac{\nabla^2\psi}{a^2}+2H\frac{\nabla^2}{a}(a\dot{E}-B)\right]f_Q\notag\\&+\left[6H^2\delta Q+6H\dot{H}\frac{\nabla^2B}{a}\right]f_{QQ}+\left[36H\dot{H}(2H\phi+\dot{\psi})-3(3H^2+\dot{H})\delta Q\right.\notag\\&\left.+6H^2\delta C+3H\dot{\delta Q}-12H\dot{H}\nabla^2\dot{E}-\frac{\nabla^2\delta Q}{a^2}-3(4H\dot{H}+\ddot{H})\frac{\nabla^2 B}{a}\right]f_{QC}\notag\\&+\left[-3(6H\dot{H}+\ddot{H})\left(6\dot{\psi}+12H\phi+\frac{1}{a}\nabla^2(B-2a\dot{E})\right)-3(3H^2+\dot{H})\delta C\right.\notag\\&\left.+3H\dot{\delta C}-\frac{\nabla^2\delta C}{a^2}\right]f_{CC}+\left[18H(6H\dot{H}+\ddot{H})\delta Q-36H^2\dot{H}\delta C\right]f_{QCC}\notag\\&-36H^2\dot{H}f_{QQC}\delta Q+18H(6H\dot{H}+\ddot{H})f_{CCC}\delta C\,,\label{fefqc11}
    %\end{align}    
    \\
    %\begin{align}
    \kappa^2 (\bar\rho+\bar{p}) u=&2(H\phi+\dot{\psi})f_Q-6H\dot{H}(\phi+3\psi+3aHB-\nabla^2E)f_{QQ}\notag\\&+\left[3(4H\dot{H}+\ddot{H})\phi+9(8H\dot{H}+\ddot{H})\psi+H\delta Q-\dot{\delta Q}\right.\notag\\&\left.+3(8H\dot{H}+\ddot{H})(3aHB-\nabla^2E)\right]f_{QC}+\left[3(6H\dot{H}+\ddot{H})(\phi-3\psi)+H\delta C\notag\right.\\&-\dot{\delta C}\left.+3(6H\dot{H}+\ddot{H})(\nabla^2E-3aHB)\right]f_{CC}\notag\\&+\left[-6(6H\dot{H}+\ddot{H})\delta Q+12H\dot{H}\delta C\right]f_{QCC}\notag\\&-6(6H\dot{H}+\ddot{H})f_{CCC}\delta C+12H\dot{H}f_{QQC}\delta Q\,,\label{fefqc22}
    \end{align}
    \\
    \begin{align}
    \kappa^2 \delta p=&\left[2(3H^2+2\dot{H})\phi+2H(\dot{\phi}+3\dot{\psi})+2\ddot{\psi}+\frac{2}{3}\frac{\nabla^2(\phi-\psi)}{a^2}-2H\nabla^2\dot{E}\right.\notag\\&\left.-\frac{2}{3}\nabla^2\ddot{E}+\frac{2}{3a}(\nabla^2\dot{B}+2H\nabla^2B)\right]f_Q-2\left[12H\dot{H}(2H\phi+\dot{\psi})\right.\notag\\&\left.-4H\dot{H}\nabla^2\dot{E}+(3H^2+\dot{H})\delta Q+H\dot{\delta Q}+H\dot{H}\frac{\nabla^2B}{a}\right]f_{QQ}\notag\\&+\left[12(6\dot{H}^2+6H\ddot{H}+\dddot{H})\phi+6(\ddot{H}+6H\dot{H})\dot{\phi}+3(3H^2+\dot{H})\delta C-\ddot{\delta C}\right.\notag\\&\left.+3(6H\dot{H}+\ddot{H})\frac{\nabla^2B}{a}+\frac{2}{3}\frac{\nabla^2\delta C}{a^2}\right]f_{CC}+\left[24\dot{H}(6H^2-\dot{H})\phi\right.\notag\\&\left.+12(\ddot{H}+6H\dot{H})\dot{\psi}-12H\dot{H}\dot{\phi}-4(\ddot{H}+6H\dot{H})\nabla^2\dot{E}\right.\notag\\&\left.+(3H^2+\dot{H})(3\delta Q-2\delta C)-2H\dot{\delta C}-\ddot{\delta Q}+\ddot{H}\frac{\nabla^2B}{a}+\frac{2}{3}\frac{\nabla^2\delta Q}{a^2}\right]f_{QC}\notag\\&-6\left[48H\dot{H}(6H\dot{H}+\ddot{H})\phi+(6\dot{H}^2+6H\ddot{H}+\dddot{H})\delta Q\right.\notag\\&\left.+2\dot{H}(6H^2-\dot{H})\delta C+2(\ddot{H}+6H\dot{H})\dot{\delta Q}-4H\dot{H}\dot{\delta C}\right]f_{QCC}\notag\\&+24H^2\dot{H}f_{QQQ}\delta Q-12\dot{H}\left[H^2(6\delta Q-2\delta C-24\dot{H}\phi)-\dot{H}\delta Q\right.\notag\\&\left.-2H\dot{\delta Q}\right]f_{QQC}+6\left[12(6H\dot{H}+\ddot{H})^2\phi-(6\dot{H}^2+6H\ddot{H}+\dddot{H})\delta C\right.\notag\\&\left.-2(6H\dot{H}+\ddot{H})\dot{\delta C}\right]f_{CCC}-36(6H\dot{H}+\ddot{H})^2f_{CCCC}\delta C\notag\\&+144H\dot{H}\left[H\dot{H}(6\delta Q-\delta C)+\ddot{H}\delta Q\right]f_{QQQC}-144H^2\dot{H}^2f_{QQQC}\delta Q\notag\\&-36(6H\dot{H}+\ddot{H})\left[2H\dot{H}(3\delta Q-2\delta C)+\ddot{H}\delta Q\right]f_{QCCC}\notag\\&\,,\label{fefqc33}
    %\end{align}
    \\
    %\begin{align}
    -\kappa^2 a^2 \Pi=0=&\left[\phi-\psi+a\dot{B}+2aHB-3a^2H\dot{E}-a^2\ddot{E}\right]f_Q-12aH\dot{H}[B-a\dot{E}]f_{QQ}\notag\\&+f_{CC}\delta C-[6a(6H\dot{H}+\ddot{H})(a\dot{E}-B)-\delta Q]f_{QC}\,,\label{fefqc44}
\end{align}
where\footnote{refer to appendix \ref{A00} equation \ref{A10}}
\begin{align}
    \delta Q=&-2H\left(-6\phi H-6\dot{\psi}+2\nabla^2\dot{E}-\frac{\nabla^2B}{2a}\right)\,,\label{dq}\\      
    \delta C=&-2\left(6\phi(3H^2+\dot{H})+3H(\dot{\phi}+6\dot{\psi})+3\ddot{\psi}+\frac{\nabla^2(\phi-2\psi)}{a^2}+7H\frac{\nabla^2B}{2a}\right.\notag\\&\left.+\frac{\nabla^2\dot{B}}{a}-6H\nabla^2\dot{E}-\nabla^2\ddot{E}\right)\,.\label{dc}
\end{align}
The set of equations (\ref{fefqc11})-(\ref{fefqc44}) are for a general $f(Q, C)$ gravity theory, from which the perturbation equations of $f(\mathring{R})$ gravity can be  retrieved as described in the appendix \ref{fqctofr}.
%Equations (\ref{fefqc11})-(\ref{fefqc44}) can be reduced to equations in $f(R)$ gravity \cite{felice2010}.
%%%%%%%%%%%%%%%%%%%%%%%%%%%%%%%%%%%%%%%%%%%%%%%%%%%%%%%%%%%%%%%%%%%%%%%%%%%%%%%%%%%%%%%%%%%%
\section{Perturbation of connection field equation and conservation equations}\label{V}
%%%%%%%%%%%%%%%%%%%%%%%%%%%%%%%%%%%%%%%%%%%%%%%%%%%%%%%%%%%%%%%%%%%%%%%%%%%%%%%%%%%%%%%%%%%%
In this section, we study the perturbation of the connection field equation (\ref{eqn:FE2-invar}) in the absence of hypermomentum tensor. In \cite{avik2023a}, De et al. introduced $f(Q, C)$ gravity and proved that the connections field equation is equivalent to continuity equation in the absence of hypermomentum tensor. Therefore, perturbation of the connection field equation (\ref{eqn:FE2-invar}) is
\begin{align}
    \kappa^2\delta(\mathring{\nabla}_\mu T^\mu{}_\nu)=&2\left[\delta(\mathring{\nabla}_\mu P^{\lambda\mu}{}_\nu)+(\delta P^{\lambda\mu}{}_\nu)\mathring{\nabla}_\mu\right]\nabla_\lambda(f_Q-f_C)\notag\\&+2\left(\mathring{\nabla}_\mu P^{\lambda\mu}{}_\nu+P^{\lambda\mu}{}_\nu \delta[\mathring{\nabla}_\mu\right)\nabla_\lambda(f_Q-f_C)]\notag\\&+\left[\delta\mathring{G}^\lambda{}_\nu+\frac{\delta Q}{2}\delta^\lambda{}_\nu\right]\nabla_\lambda(f_Q-f_C)+\left(\mathring{G}^\lambda{}_\nu+\frac{Q}{2}\delta^\lambda{}_\nu\right)\delta\left[\nabla_\lambda(f_Q-f_C)\right]\,.\label{coneq}
\end{align}
From the fundamental law of conservation, we let $\delta(\mathring{\nabla}_\mu T^\mu{}_\nu)=0$, thence from equation (\ref{coneq}) the perturbation of energy and momentum conservation equation can be expressed as
\begin{align}
    G_1f_{QQ}+G_2f_{QC}+G_3f_{CC}+G_4f_{QQQ}+G_5f_{QQC}+G_6f_{QCC}+G_7f_{CCC}=&0\,,\label{eceq}\\
    G_{11}f_{QQ}+G_{22}f_{QC}+G_{33}f_{CC}+G_{44}f_{QQQ}+G_{55}f_{QQC}+G_{66}f_{QCC}+G_{77}f_{CCC}=&0\,,\label{mceq}
\end{align}
where
\begin{align}
    G_1=&6H^2\dot{\delta Q}+6H^2\dot{H}\left(6H\phi-3(\dot{\phi}-2\dot{\psi})-2\nabla^2\dot{E}\right)-\frac{H}{2}\frac{\nabla^2\delta Q}{a^2}\notag\\&-6(\dot{H}^2+H\ddot{H}-H^2\dot{H})\frac{\nabla^2B}{a}\,,\\
    G_2=&6H^2\dot{\delta C}+3\ddot{H}\frac{\nabla^2\dot{B}}{a}+3(8\dot{H}^2+5H\ddot{H}+\dddot{H})\frac{\nabla^2B}{a}+\frac{H}{2}\frac{\nabla^2(\delta Q-\delta C)}{a^2}\notag\\&+H\dot{H}\left(-504H^2\phi-504H\dot{\psi}+24\frac{\nabla^2(\phi+2\psi)}{a^2}+6\delta Q-48H\frac{\nabla^2B}{a}\right.\notag\\&\left.+24\frac{\nabla^2\dot{B}}{a}-24\frac{\nabla^2\nabla^2E}{a^2}+168H\nabla^2\dot{E}\right)\notag\\&+3\ddot{H}\left(-24H^2\phi-24H\dot{\psi}+\frac{\nabla^2(\phi+3\psi)}{a^2}-\frac{\nabla^2\nabla^2E}{a^2}+8H\nabla^2\dot{E}\right)\,,\\
    G_3=&3(6H\dot{H}+\ddot{H})\left(12H^2\phi+12H\dot{\psi}-\frac{\nabla^2(\phi-\psi)}{a^2}-\delta Q-H\frac{\nabla^2B}{a}-\frac{\nabla^2\dot{B}}{a}\right.\notag\\&\left.+\frac{\nabla^2\nabla^2E}{a^2}-4H\nabla^2\dot{E}\right)+\frac{H}{2}\frac{\nabla^2\delta C}{a^2}-3(6\dot{H}^2+6H\ddot{H}+\dddot{H})\frac{\nabla^2B}{a}\,,
    %\end{align}
    \\
    %\begin{align}
    G_4=&-72H^3\dot{H}\delta Q+72H^2\dot{H}\frac{\nabla^2B}{a}\,,\\
    G_5=&36\left[(6H^3\dot{H}+H^2\ddot{H})\delta Q-2H^3\dot{H}\delta C-2H\dot{H}\left(7H\dot{H}+\ddot{H}\right)\frac{\nabla^2B}{a}\right]\,,
    %\end{align}
    \\
    %\begin{align}
    G_6=&18\left[2H^2(6H\dot{H}+\ddot{H})\delta C+(60H^2\dot{H}^2+16H\dot{H}\ddot{H}+\ddot{H}^2)\frac{\nabla^2B}{a}\right]\,,
    %\end{align}
    \\
    %\begin{align}
    G_7=&-18(36H^2\dot{H}^2+12H\dot{H}\ddot{H}+\ddot{H}^2)\frac{\nabla^2B}{a}\,,
%    \\
 \end{align}
% and
 \begin{align}
G_{11}=&\left(6H^3\dot{H}+6(\dot{H}^2+H\ddot{H})\right)\phi+\left(18(\dot{H}^2+H\ddot{H})+108H^3\dot{H}\right)\psi\notag\\&+\left(12H^4\dot{H}-6H^5\dot{H}+12H\dot{H}\frac{\nabla^2}{a^2}\right)aB+\left(6H^3\dot{H}-12aH^3\dot{H}\right)\dot{B}\notag\\&+\left(6(\dot{H}^2+H\ddot{H})+96H^3\dot{H}+\frac{2}{3}H^2\right)\nabla^2E+\frac{4}{3}\nabla^2\dot{E}\notag\\&+(3H^4+2H\dot{H})\frac{\delta Q}{2}-3H^2\dot{\delta Q}\,,\\
    G_{22}=&\left(-9H^2(8H\dot{H}+\ddot{H})-6(4\dot{H}^2+4H\ddot{H}+3\dddot{H})+24\frac{H^2\dot{H}}{a^2}\right)\phi\notag\\&-\left(48H^2(8H\dot{H}+\ddot{H})+18(4\dot{H}^2+4H\ddot{H}+3\dddot{H})\right)\psi-6(8H\dot{H}+\ddot{H})\dot{\phi}\notag\\&+\left(24\frac{H\dot{H}}{a^2}+18(8H\dot{H}+\ddot{H})\right)\dot{\psi}-\left(48\frac{H^2\dot{H}^2}{a^2}+6(8H\dot{H}+\ddot{H})\frac{\nabla^2}{a}\right)B\notag\\&+2(8H\dot{H}+\ddot{H})\nabla^2\dot{E}+(6H^2+3H^4+6H\dot{H})\frac{\delta Q}{2}-(3H^4+2H\dot{H})\frac{\delta C}{2}\notag\\&+\frac{3}{2}H^2(\dot{\delta Q}-\dot{\delta C})\,,
    \end{align}
\begin{align}
 G_{33}=&\left(-3H^2(6H\dot{H}+\ddot{H})\left(1+\frac{4}{a^2}\right)+3(6\dot{H}^2+6H\ddot{H}+\dddot{H})\right)\phi-3(6H\dot{H}+\ddot{H})\dot{\phi}\notag\\&+\left(27H^2(6H\dot{H}+\ddot{H})+9(6\dot{H}^2+6H\ddot{H}+\dddot{H})\right)\left(\psi-\frac{1}{3}\nabla^2E\right)\notag\\&-3(6H\dot{H}+\ddot{H})\left(\frac{4}{a}-1\right)\dot{\psi}-\left(6H^4(6H\dot{H}+\ddot{H})-3(6H\dot{H}+\ddot{H})\frac{\nabla^2}{a^2}\right)aB\notag\\&+6(6H\dot{H}+\ddot{H})\left(H^2+4\frac{H\dot{H}}{a^2}\right)a\dot{B}+(6H\dot{H}+\ddot{H})\left(\frac{4}{a}-\frac{2}{a^2}+3\right)\nabla^2\dot{E}\notag\\&+\left(6H^2+3H^4+6H\dot{H}\right)\frac{\delta C}{2}+\frac{3}{2}H^2\dot{\delta C}\,,\\
    G_{44}=&72H^2\dot{H}\left(-\phi-\psi+\frac{1}{3}\nabla^2E\right)+18H^3\dot{H}\,,\\
    G_{55}=&72H\dot{H}\left(\phi+3\psi-\nabla^2E\right)-9H^2(8H\dot{H}+\ddot{H})\delta Q+18H^3\dot{H}\delta C\,,\\
    G_{66}=&18(10H\dot{H}+\ddot{H})(6H\dot{H}+\ddot{H})\left(-\phi-3\psi+\nabla^2E\right)+9H^2(6H\dot{H}+\ddot{H})\delta Q\notag\\&-9H^2(8H\dot{H}+\ddot{H})\delta C\,,\\
    G_{77}=&18(6H\dot{H}+\ddot{H})^2(\phi+3\psi-\nabla^2E)+9H^2(6H\dot{H}+\ddot{H})\delta C\,.
\end{align}
According to the law of conservation, the perturbations of the conservation equation $\delta(\mathring{\nabla}^\mu T_{\mu\nu})=0$ for a perfect fluid gives us 
\begin{align}
    \dot{\delta}&=3H(\omega-c_s{}^2)\delta+(1+\omega)\left[\frac{\nabla^2}{a^2}(u+aB)+3\dot{\psi}-\nabla^2\dot{E}\right]\,,\label{ddel}\\
    \dot{u}&=\frac{c_s{}^2}{1+\omega}\delta+3H\omega u+\phi\,,\label{dv}
\end{align}
for the time and spatial components respectively. Furthermore, equations (\ref{ddel}) and (\ref{dv}) can be combined together by taking time derivative of equation (\ref{ddel}) and substituting equation (\ref{dv}) to produce
\begin{align}\label{add1}
    \ddot{\delta}=&\left[3(2\omega-c_s{}^2)-2\right]H\dot{\delta}+3(\omega-c_s{}^2)\left[\dot{H}+H^2(2-3\omega)+\frac{c_s{}^2}{3(\omega-c_s{}^2)}\frac{\nabla^2}{a^2}\right]\delta\notag\\&+(1+\omega)\left[3(2-3\omega)H\dot{\psi}+3\ddot{\psi}+\frac{\nabla^2\phi}{a^2}+(-2+3\omega)H\nabla^2\dot{E}-\nabla^2\ddot{E}\right.\notag\\&\left.+(1-3\omega)H\frac{\nabla^2B}{a}+\frac{\nabla^2\dot{B}}{a}\right]\,.
\end{align}
For a matter-dominated universe, radiation density is ignored compared to matter density, so that we can set $c_s{}^2\equiv\frac{\delta P}{\delta \rho}\approxeq\frac{\bar p}{\bar \rho}=\omega=0$ and the equation (\ref{add1}) above reduces to 
\begin{equation}\label{denp11}
\ddot{\delta}^m=-2H\dot{\delta}^m+6H\dot{\psi}+3\ddot{\psi}+\frac{\nabla^2}{a^2}\left[\phi-2a^2H\dot{E}-a^2\ddot{E}+a(HB+\dot{B})\right]\,.
\end{equation}

%%%%%%%%%%%%%%%%%%%%%%%%%%%%%%%%%%%%%%%%%%%%%%%%%%%%%%%%%%%%%%%%%%%%%%%%%%%%
\subsection{Quasi-static and sub-horizon approximation}\label{sQS}
%%%%%%%%%%%%%%%%%%%%%%%%%%%%%%%%%%%%%%%%%%%%%%%%%%%%%%%%%%%%%%%%%%%%%%%%%%%%
In the following, we analyse the differential equations governing the dynamics of the density perturbation by applying some approximation process broadly used in cosmological perturbative analysis. We also use the minimally coupled models $f(Q,C)=f_1(Q)+f_2(C)$ for the following computations. Two approximations are widely used in modified gravity to demonstrate structure formations in the universe, namely, the quasi-static (QS) approximation which states that the time derivative of the potentials are negligible and sub-horizon (SH) approximation which states that the wavemodes are $k>>a H$ \cite{quintana2023}, we extend these assumption to $\dot{H}<<\frac{k^2}{a^2}, \ddot{H}<<H\frac{k^2}{a^2},\dddot{H}<<H^2\frac{k^2}{a^2}$ etc. \cite{gorbunov2011} and name it strong SH approximation.
% {\color{brown}{where the Hubble parameters and its derivatives are within the observable wavelength in this approximation}}. 
\par
The approximation technique is regularly used in modified gravity theories to attain solvable equations. In practice, it is assumed that the time derivatives of gravitational potentials are subdominant with respect to the spatial derivatives. %Such an approximation produces reasonably good results within the Hubble radius. 
As described recently \cite{quintana2023}, the sub-horizon approximation can be safely used for scales $0.06 \,\text{h/Mpc} \lessapprox k \lessapprox0.2 \, \text{h/Mpc}$ and in such scale, the quasi-static approximation provides a significantly good description of the late-time cosmological dynamics. In GR this approximation was first used by Starobinsky in the presence of a minimally coupled scalar field \cite{567}, which was numerically confirmed in \cite{403}. This was further extended to scalar-tensor theories \cite{93, 171}, the $f(\mathring{R})$ gravity \cite{tsujikawa2007, bean2007, dombriz2008, girones2010, motohashi2011, tsujikawa2008, chiu2015, eingorn2014}, the $f(Q)$ theory \cite{jimenez2020}, among others (refer to \cite{saridakis2021} for more details). 
\par
We use the Fourier transform of the form
\begin{align}
    \phi(\vec{x},t)=\int\frac{d^3\vec{k}}{\sqrt{2\pi}^3}\tilde{\phi}_{\vec{k}}(t) e^{i\vec{k}\cdot \vec{x}}\,; \ \ \ \  \psi(\vec{x},t)=\int\frac{d^3\vec{k}}{\sqrt{2\pi}^3}\tilde{\psi}_{\vec{k}}(t) e^{i\vec{k}\cdot \vec{x}}\,,
\end{align}
and $\nabla^2\equiv -k^2$. Therefore, equations (\ref{fefqc11}) and (\ref{fefqc44}) reduce to
\begin{align}
    \kappa^2\bar\rho \tilde{\delta}_{\vec{k}}=&\left[-6H^2\tilde{\phi}_{\vec{k}}-2\frac{k^2}{a^2}\tilde{\psi}_{\vec{k}}+2H\frac{k^2}{a}\tilde{B}_{\vec{k}}\right]f_Q\notag\\&+
    \left[72H^4\tilde{\phi}_{\vec{k}}-6H(H^2+\dot{H})\frac{k^2}{a}\tilde{B}_{\vec{k}}\right]f_{QQ}\notag\\&+2\frac{k^2}{a^2}\left[\frac{k^2(\tilde{\phi}_{\vec{k}}-2\tilde{\psi}_{\vec{k}})}{a^2}+7H\frac{k^2\tilde{B}_{\vec{k}}}{2a}\right]f_{CC}\,,\label{shqsdelta}\\
    2f_{CC}\left[-\frac{k^2(\tilde{\phi}_{\vec{k}}-2\tilde{\psi}_{\vec{k}})}{a^2}-7H\frac{k^2\tilde{B}_{\vec{k}}}{2a}\right]=&\left[\tilde{\phi}_{\vec{k}}-\tilde{\psi}_{\vec{k}}+2aH\tilde{B}_{\vec{k}}\right]f_Q-12aH\dot{H}\tilde{B}_{\vec{k}}f_{QQ}\,,\label{shqsij}
\end{align}
where we use the perturbation of non-metricity scalar (\ref{dq}) and boundary term  (\ref{dc}) with QS and strong SH approximations,
\begin{equation}
    \delta Q=-2H\left(-6H\tilde{\phi}_{\vec{k}}+\frac{k^2\tilde{B}_{\vec{k}}}{2a}\right)\,,\qquad\qquad     
    \delta C=2\left(\frac{k^2(\tilde{\phi}_{\vec{k}}-2\tilde{\psi}_{\vec{k}})}{a^2}+7H\frac{k^2\tilde{B}_{\vec{k}}}{2a}\right)\,.
\end{equation}

Furthermore, in this model the energy conservation equation (\ref{eceq}) can be expressed as
\begin{equation}
    G_1f_{QQ}+G_3f_{CC}+G_4f_{QQQ}+G_7f_{CCC}=0\,,\label{eceq1}
\end{equation}
with
\begin{align}
    G_1=&6H^3\frac{k^2\tilde{\phi}_{\vec{k}}}{a}-H^2\frac{k^2k^2\tilde{B}_{\vec{k}}}{2a^3}\,,\qquad\qquad\qquad
    G_3=H\frac{k^2}{a^2}\left[\frac{k^2(2\tilde{\psi}_{\vec{k}}-\tilde{\phi}_{\vec{k}})}{a^2}-7H\frac{k^2\tilde{B}_{\vec{k}}}{2a}\right]\,,\label{eqg1}\\
    G_4=&\dot{Q}\left[72H^4\tilde{\phi}_{\vec{k}}-6H(H^2-1)\frac{k^2\tilde{B}_{\vec{k}}}{a}\right]\,,\qquad
    G_7=\frac{\dot{C}^2}{2}\frac{k^2\tilde{B}_{\vec{k}}}{a}\,.\label{eqg2}
\end{align}
Using the relations $\dot{Q}f_{QQQ}=\dot{f}_{QQ}$, $\dot{C}f_{CCC}=\dot{f}_{CC}$ and the approximation $\dot{f}_{QQ}\ll f_{QQ}$, the last two terms on the LHS of equation (\ref{eceq1}) are negligible. Therefore, equation (\ref{eceq1}) can be written as follows using equation (\ref{eqg1}):
\begin{equation}
    \left(6H^3\frac{k^2\tilde{\phi}_{\vec{k}}}{a}+H^2\frac{k^2k^2\tilde{B}_{\vec{k}}}{2a^3}\right)f_{QQ}+H\frac{k^2}{a^2}\left[\frac{k^2(2\tilde{\psi}_{\vec{k}}-\tilde{\phi}_{\vec{k}})}{a^2}-7H\frac{k^2\tilde{B}_{\vec{k}}}{2a}\right]f_{CC}=0\,,
\end{equation}
from which we obtain
\begin{equation} 
     \frac{k^2\tilde{B}_{\vec{k}}}{a}=\frac{12H^3\tilde{\phi}_{\vec{k}} f_{QQ}-2Hf_{CC}\frac{k^2(\tilde{\phi}_{\vec{k}}-2\tilde{\psi}_{\vec{k}})}{a^2}}{H^2(f_{QQ}+7f_{CC})}\,.\label{nablasqrB}
\end{equation}
Substituting this in  equations (\ref{shqsdelta}) and (\ref{shqsij}), we get
\begin{align}
    \kappa^2\bar\rho(f_{QQ}+7f_{CC})\delta=&\left[\left(18H^2f_{QQ}\frac{a^2}{k^2}-4f_{CC}\right)f_Q+2f_{QQ}f_{CC}\frac{k^2}{a^2}\right.\notag\\&\left.-72H^2\dot{H}f_{QQ}{}^2\frac{a^2}{k^2}\right]\frac{k^2}{a^2}\tilde{\phi}_{\vec{k}}\notag\\&-\left[2(f_{QQ}+3f_{CC})f_Q+4f_{QQ}f_{CC}\frac{k^2}{a^2}\right]\frac{k^2}{a^2}\tilde{\psi}_{\vec{k}}\,,\label{dro}\\
    \frac{k^2}{a^2}\tilde{\psi}_{\vec{k}}=&\left(\frac{f_Q(f_{QQ}+3f_{CC})+2f_{QQ}f_{CC}\frac{k^2}{a^2}}{f_Q(f_{QQ}-f_{CC})+4f_{QQ}f_{CC}\frac{k^2}{a^2}}\right)\frac{k^2}{a^2}\tilde{\phi}_{\vec{k}}\,.\label{si}
    % \frac{k^2}{a^2}\tilde{\psi}_{\vec{k}}=&\left(\frac{{\bf{\left(f_Q+f_{CC}\frac{k^2}{a^2}\right)f_{QQ}}}+3f_Qf_{CC}}{f_Qf_{QQ}-f_Qf_{CC}+2f_{QQ}f_{CC}\frac{k^2}{a^2}}\right)\frac{k^2}{a^2}\tilde{\phi}_{\vec{k}}\,.\label{si}
\end{align}
By using equation (\ref{si}) in (\ref{dro}) and rearrange to get
\begin{align}
    \frac{k^2}{a^2}\tilde{\phi}=&\kappa^2\bar\rho(f_{QQ}+7f_{CC})\left[\left(18H^2f_{QQ}\frac{a^2}{k^2}-4f_{CC}\right)f_Q+2f_{QQ}f_{CC}\frac{k^2}{a^2}\right.\notag\\&\left.-72H^2\dot{H}f_{QQ}{}^2\frac{a^2}{k^2}-\left(2(f_{QQ}+3f_{CC})f_Q+4f_{QQ}f_{CC}\frac{k^2}{a^2}\right)\right.\notag\\&\left.\times\left(\frac{f_Q(f_{QQ}+3f_{CC})+2f_{QQ}f_{CC}\frac{k^2}{a^2}}{f_Q(f_{QQ}-f_{CC})+4f_{QQ}f_{CC}\frac{k^2}{a^2}}\right)\right]^{-1}\delta\,.\label{phi1}
\end{align}
On the other hand, using equation (\ref{si}) in equation (\ref{nablasqrB}), we obtain
\begin{align}
    \frac{k^2}{a}\tilde{B}_{\vec{k}}=&\frac{1}{H^2(f_{QQ}+7f_{CC})}\left[12H^3f_{QQ}\frac{a^2}{k^2}-2Hf_{CC}\right.\notag\\&\left.+4Hf_{CC}\left(\frac{f_Q(f_{QQ}+3f_{CC})+2f_{QQ}f_{CC}\frac{k^2}{a^2}}{f_Q(f_{QQ}-f_{CC})+4f_{QQ}f_{CC}\frac{k^2}{a^2}}\right)\right]\frac{k^2}{a^2}\tilde{\phi}_{\vec{k}}\,.\label{nablasqrb1}
\end{align}
Using QS approximation, equation (\ref{denp11}) can be expressed as
\begin{equation}\label{ddotdell}
    \ddot{\tilde{\delta}}^m{}_{\vec{k}}=-2H\dot{\tilde{\delta}}^m{}_{\vec{k}}-\frac{k^2}{a^2}\left[\tilde{\phi}_{\vec{k}}+aH\tilde{B}_{\vec{k}}\right]\,.
\end{equation}
By using equation (\ref{phi1}) in matter era and equation (\ref{nablasqrb1}), the expression $\frac{k^2}{a^2}\left[\tilde{\phi}_{\vec{k}}+aH\tilde{B}_{\vec{k}}\right]$ in equation (\ref{ddotdell}) can be written as
\begin{align}
    \frac{k^2}{a^2}(\tilde{\phi}_{\vec{k}}+aH\tilde{B}_{\vec{k}})=&\frac{\kappa^2}{2}\bar\rho^m \left(f_{QQ}+5f_{CC}+4f_{CC}\left(\frac{f_Q(f_{QQ}+3f_{CC})+2f_{QQ}f_{CC}\frac{k^2}{a^2}}{f_Q(f_{QQ}-f_{CC})+4f_{QQ}f_{CC}\frac{k^2}{a^2}}\right)\right)\notag\\&\times\left[f_Q\left(9H^2f_{QQ}\frac{a^2}{k^2}-2f_{CC}\right)+f_{QQ}f_{CC}\frac{k^2}{a^2}-36H^2\dot{H}f_{QQ}{}^2\frac{a^2}{k^2}\right.\notag\\&\left.-\left((f_{QQ}+3f_{CC})f_Q+2f_{QQ}f_{CC}\frac{k^2}{a^2}\right)\right.\notag\\&\left.\times\left(\frac{f_Q(f_{QQ}+3f_{CC})+2f_{QQ}f_{CC}\frac{k^2}{a^2}}{f_Q(f_{QQ}-f_{CC})+4f_{QQ}f_{CC}\frac{k^2}{a^2}}\right)\right]^{-1}\tilde{\delta}^m{}_{\vec{k}}\,.\label{kveceqn}
\end{align}
Therefore, equation (\ref{ddotdell}) can be expressed as
\begin{equation}
    \ddot{\tilde{\delta}}^m{}_{\vec{k}}+2H\dot{\tilde{\delta}}^m{}_{\vec{k}}-4\pi G_{eff}\bar \rho^m\tilde{\delta}^m{}_{\vec{k}}=0\,,\label{ddotdelf}
\end{equation}
where we have used $\kappa^2=8\pi G$ and
\begin{align}
    G_{eff}=&-G\left(f_{QQ}+5f_{CC}+4f_{CC}\left(\frac{f_Q(f_{QQ}+3f_{CC})+2f_{QQ}f_{CC}\frac{k^2}{a^2}}{f_Q(f_{QQ}-f_{CC})+4f_{QQ}f_{CC}\frac{k^2}{a^2}}\right)\right)\notag\\&\times\left[f_Q\left(9H^2f_{QQ}\frac{a^2}{k^2}-2f_{CC}\right)+f_{QQ}f_{CC}\frac{k^2}{a^2}-36H^2\dot{H}f_{QQ}{}^2\frac{a^2}{k^2}\right.\notag\\&\left.-\left((f_{QQ}+3f_{CC})f_Q+2f_{QQ}f_{CC}\frac{k^2}{a^2}\right)\left(\frac{f_Q(f_{QQ}+3f_{CC})+2f_{QQ}f_{CC}\frac{k^2}{a^2}}{f_Q(f_{QQ}-f_{CC})+4f_{QQ}f_{CC}\frac{k^2}{a^2}}\right)\right]^{-1}\,.\label{geff1}
\end{align}
\par
\begin{figure}[!tbp]
    \centering
 \begin{minipage}[b]{0.45\textwidth}
    \includegraphics[width=\textwidth]{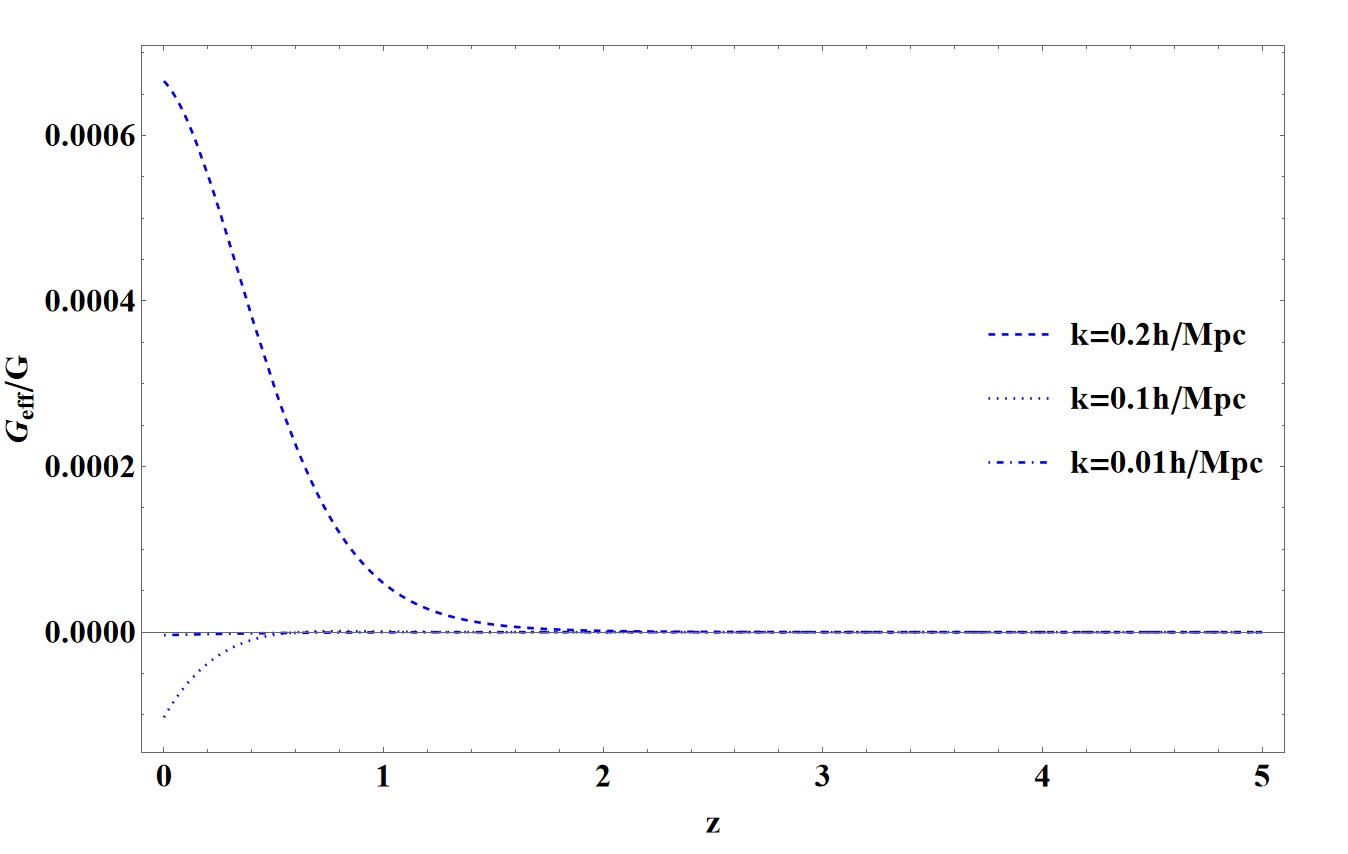}
    \caption{$G_{eff}/G$ vs redshift $z$ for $f(Q,C)=Q+\beta Q^2+\sigma C^2$}
    \label{figG1}
 \end{minipage}
 \hfill
 \begin{minipage}[b]{0.45\textwidth}
    \includegraphics[width=\textwidth]{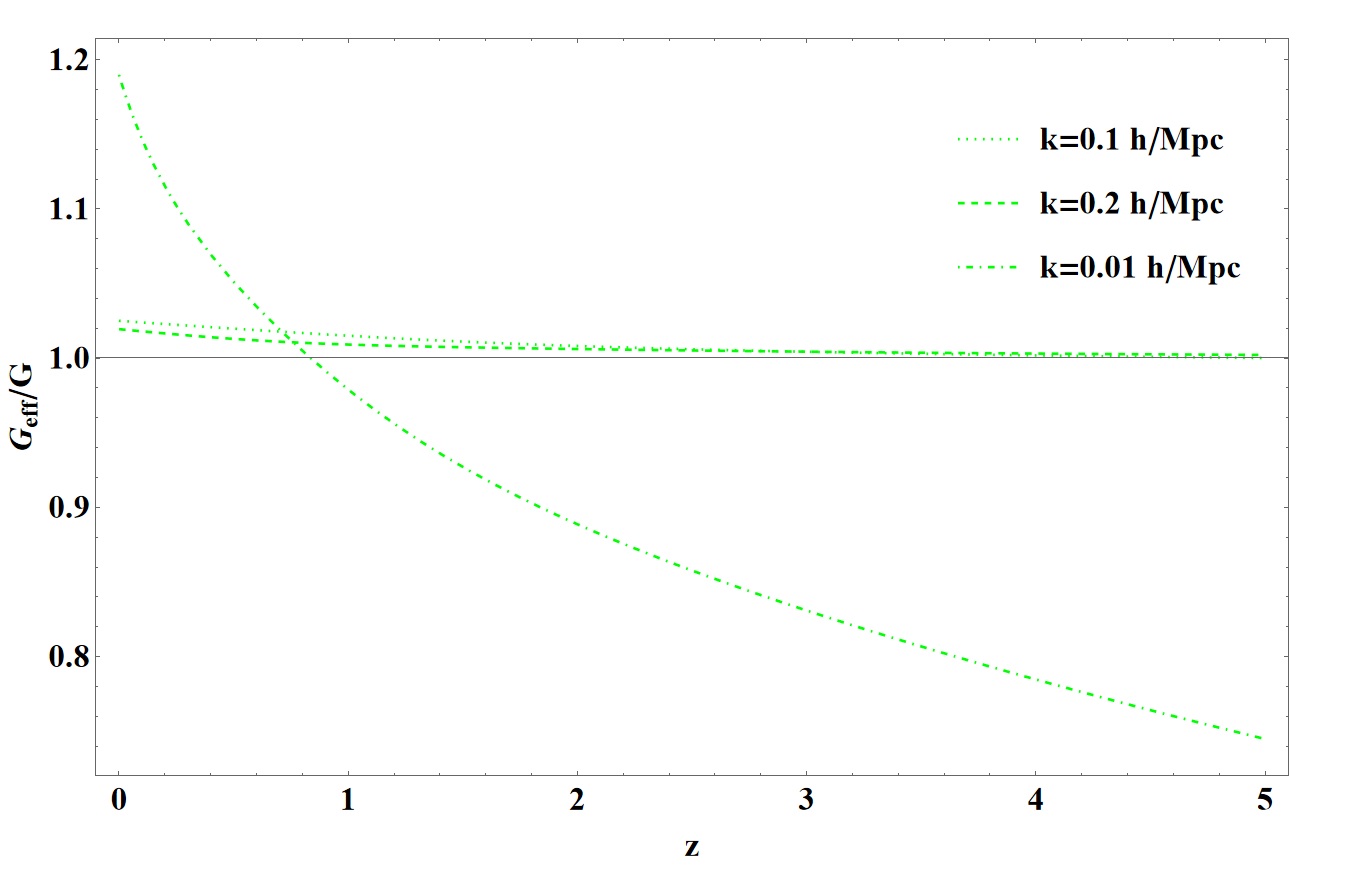}
    \caption{$G_{eff}/G$ vs redshift $z$ for $f(Q,C)=Q+\lambda\sqrt{-Q}+\Xi C^2$}
    \label{figG2}
 \end{minipage}
\end{figure}
\begin{figure}[h!]
    \centering
    \includegraphics[width=1\linewidth]{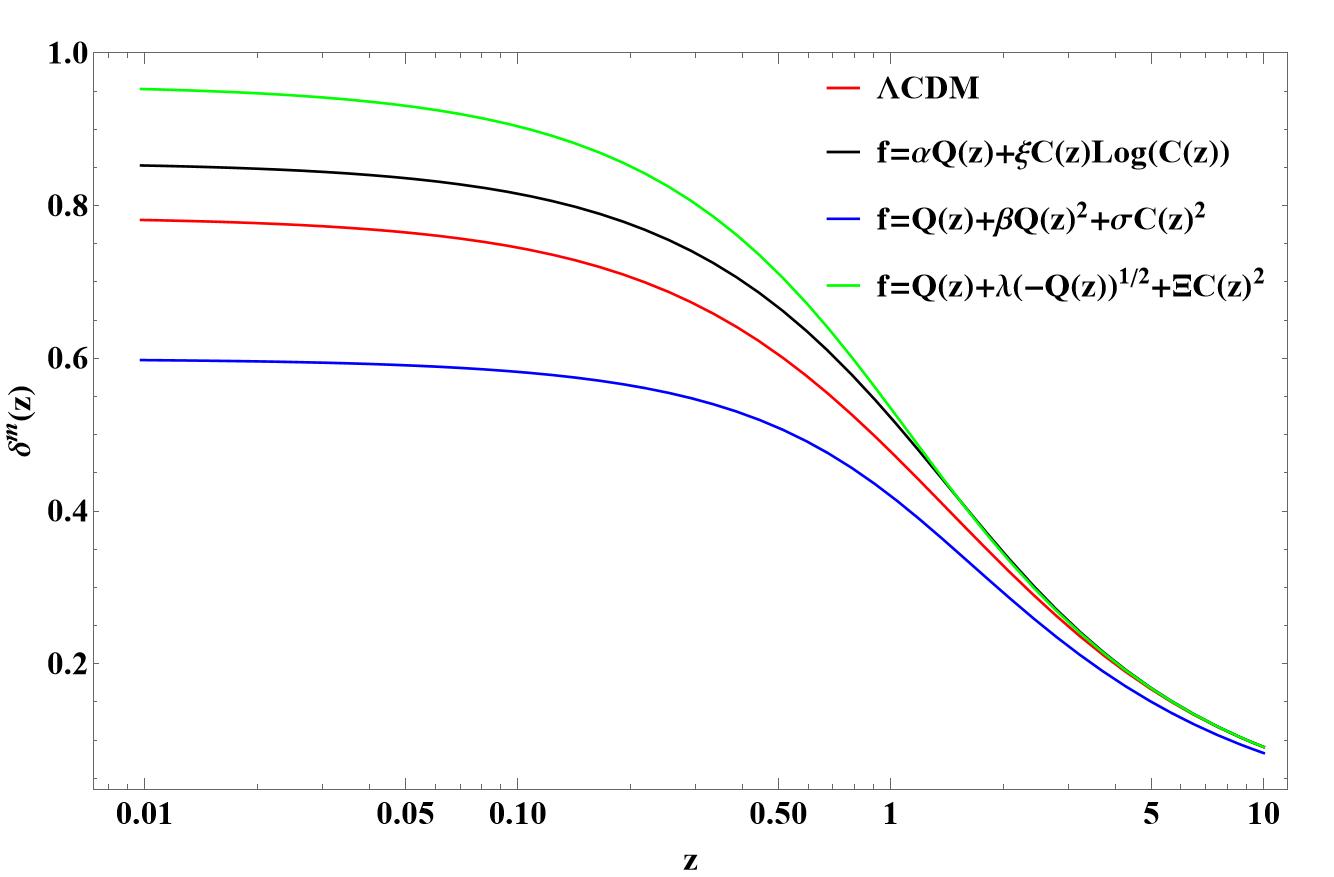}
    \caption{Matter density contrast $\delta^m$ vs redshift $z$ for several $f(Q, C)$ models
    with $k=0.01 h/Mpc$.}
    \label{fig1}
\end{figure}

As discussed in the introduction, the effective gravitational constant $G_{eff}/G$ measures the deviation of modified gravity from the standard theory of gravity. To depict this with a comparison to $\Lambda$CDM, we consider a few $f(Q,C)$ models, e.g., i) $\alpha Q+\xi C\log(c_0C)$ where $c_0=1(\text{km/s/Mpc})^2$, $\alpha=1.3$ and $\xi=-2.4$ \cite{kadam2023}, ii) $Q+\beta Q^2+\sigma C^2$ where $\beta=0.01 (\text{km/s/Mpc})^2$ and $\sigma=-0.1 (\text{km/s/Mpc})^2$ and iii) $Q+\lambda\sqrt{-Q}+\Xi C^2$ where $\lambda=1 (\text{km/s/Mpc})^{-1}$ and $\Xi=1 (\text{km/s/Mpc})^2$. For the first model, due to linear $Q$, $G_{eff}/G$ reduces to a constant $\alpha^{-1}$, the coefficient of $Q$. In Figure \ref{figG1} and Figure \ref{figG2} we plot the $G_{eff}/G$ for the other two $f(Q,C)$ models and also few variations of $k$ is considered. We plot the matter density contrast $\delta^m$ vs redshift $z$ in Figure \ref{fig1}. We consider the initial conditions $\delta(z)=\frac{1}{1+z}$ and $\delta'(z)=-\frac{1}{(1+z)^2}$ at $z=1100$.
% From Figure \ref{fig1}, we observed that the models has similar behaviour and is close to the matter density evolution for the $\Lambda$CDM model in the past. 
% On the other hand, the model deviates away from $\Lambda$CDM at present. 
% The model $f(Q,C)=\alpha Q(z)+\xi C(z)\log(c_0C(z))$ having matter density contrast of $\delta^m(z=0)=0.85$ while for $\Lambda$CDM the matter density contrast is $\delta^m{}_{\Lambda CDM}(z=0)=0.78$. For the model $f(Q,C)=Q+\lambda\sqrt{-Q}+\Xi C^2$, the $\delta^m(z=0)=0.95$ which is much away from $\Lambda$CDM. On the other hand, for the model $f(Q,C)=Q+\beta Q^2+\sigma C^2$, $\delta^m(z=0)=0.6$. Note that the values estimated here are from the theoretical computation and the obtained plots, might differ from observational values.

%%%%%%%%%%%%%%%%%%%%%%%%%%%%%%%%%%%%%%%%%%%%%%%%%%%%%%%%%%%%%%%%%%%%%%%%%%%%%%%%%%%%%%%%%%%%%%%%%%%%%%%%%%%%%%%%%%%%%%%%%%%%%%%%%%%%
\section{Density growth and growth index}\label{VI}
%%%%%%%%%%%%%%%%%%%%%%%%%%%%%%%%%%%%%%%%%%%%%%%%%%%%%%%%%%%%%%%%%%%%%%%%%%%%%%%%%%%%%%%%%%%%%%%%%%%%%%%%%%%%%%%%%%%%%%%%%%%%%%%%%%%%%%%%%%%%%%%%%%%%%%%%%%%%%%%%%%%%%
In this section, we investigate the evolution of the {\bf{matter growth rate}} $f_g$. The matter growth rate $f_g$ is defined as \cite{peebles1993, linder2007, qing2014}
\begin{equation}
    f_g\equiv\frac{d\ln{\tilde{\delta}^m}{}_{\vec{k}}}{d\ln{a}}\,.
\end{equation}

\begin{figure}[!h]
    \centering
    \includegraphics[width=\linewidth]{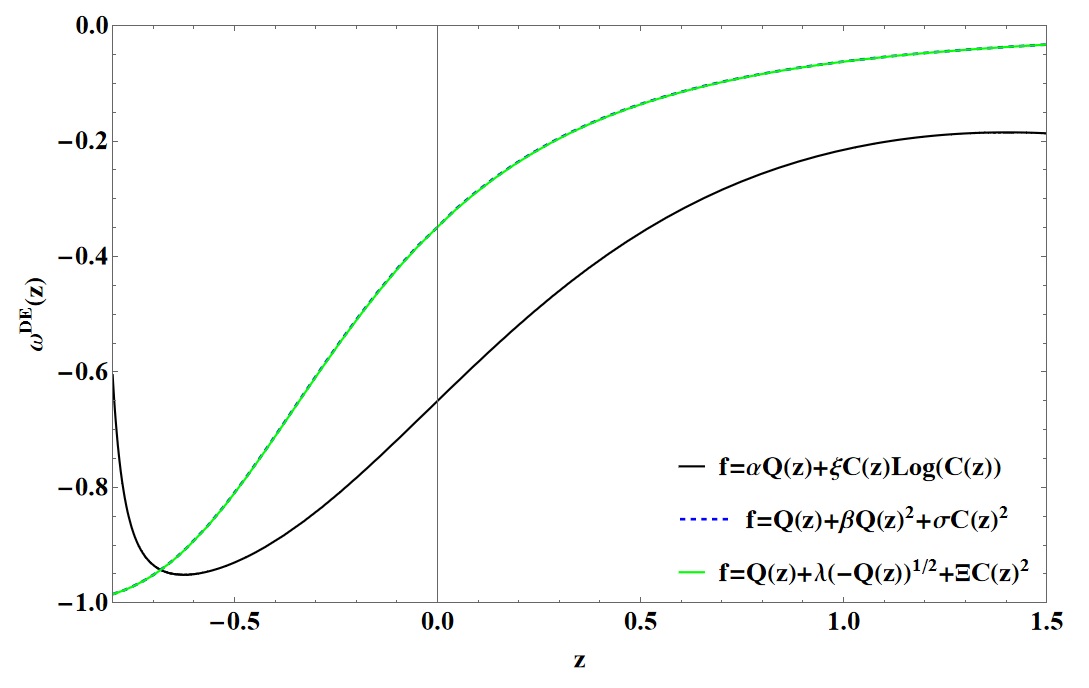}
    \caption{Dark energy equation of state parameter $\omega^{DE}$ vs redshift $z$}
    \label{fig2}
\end{figure}

Consider a matter-dominated era of the universe ($\bar p^m=0$), using equation (\ref{fle2}) and the relation $\bar p^{DE}=\omega^{DE}\bar \rho^{DE}$, we found
\begin{equation}
    -\frac{2\dot{H}}{3H^2}=\omega^{DE}(1-\Omega^{m})+1\,,\label{eeq1}
\end{equation}
where 
\begin{equation}
    \Omega^m=\frac{\Omega^{m0}}{a^3}\left(\frac{H_0}{H}\right)^2\,,\label{om}
\end{equation}
is the matter density parameter with $\Omega^{m0}$ and $H_0$ are the present value of the matter density parameter and Hubble parameter respectively. The dark energy equation of state parameter $\omega^{DE}$ is derived from equation (\ref{rode}) and (\ref{pde}) and plotted in Figure \ref{fig2} below
\begin{equation}
    \omega^{DE}=\frac{\bar p^{DE}}{\bar \rho^{DE}}=-1+\frac{-2\dot{H}(1-f_{Q})+2H\dot{f}_Q-3H\dot{f}_C+\ddot{f}_{C}}{3H^2(1-2f_Q)-\frac{f}{2}+3(3H^2+\dot{H})f_C-3H\dot{f}_C}\,.\label{wde}
\end{equation}
%{\bf{We plot Figure \ref{fig2}, the evolution of dark energy equation of state parameter $\omega^{DE}$ for the Hubble parameter $H(z)$ obtain from  equations (\ref{fle1}) and (\ref{fle2}). It is observed that the dark energy equation of state parameter $\omega^{DE}(z=0)$ for the model $f(Q, C)=\alpha Q(z)+\xi C(z)\log(c_0C(z))$ is $-0.65$ while for the models $f(Q, C)=Q+\beta Q^2+\sigma C^2$ and $f(Q, C)=Q+\lambda\sqrt{-Q}+\Xi C^2$ are $-0.35$ at present ($z=0$). From Figure (\ref{fig2}), it is observed that the $\omega^{DE}$ approaches $-1$ in the future for the models $f(Q, C)=Q+\beta Q^2+\sigma C^2$ and $f(Q,C)=Q+\lambda\sqrt{-Q}+\Xi C^2$. Apart from that, for the model $f(Q,C)=\alpha Q(z)+\xi C(z)\log(c_0C(z))$, the $\omega^{DE}(z=-0.65)=-0.95$ which is close to $-1$ and thereafter start increasing towards $0$.}}

Furthermore, converting the time derivatives in equation (\ref{eeq1}) to derivative with respect to $\log{a}$ using $\frac{d}{dt}=H\frac{d}{d\log{a}}$, we obtained
\begin{equation}
    \frac{d\log{H}}{d\log{a}}=-\frac{3}{2}[\omega^{DE}(1-\Omega^m)+1]\,.\label{dlnh}
\end{equation}
Similarly, equation (\ref{ddotdelf}) can be converted to derivative with respect to $\log{a}$ and we found
\begin{equation}
    \frac{d f_g}{d\log a}+f_g{}^2+\frac{1}{2}\left[1-3\omega^{DE}(1-\Omega^m)\right]f_g=\frac{3}{2}\left(\frac{G_{eff}}{G}\right)\Omega^m\,,\label{eveq2}
\end{equation}
where we use the relation $\kappa^2 \bar \rho^m=3H_0{}^2\Omega^{m0}a^{-3}$. Figure \ref{fig3} displays its behaviour in comparison with the $\Lambda$CDM.
{\bf{
% Since, we plot matter density contrast $\delta^m(z)$ vs redshift $z$ in Figure \ref{fig1}, we can numerically plot growth rate $f_g$ vs redshift $z$ as in Figure \ref{fig3} without solving the equation (\ref{eveq2}).
\begin{figure}[h!]
    \centering
    \includegraphics[width=\linewidth]{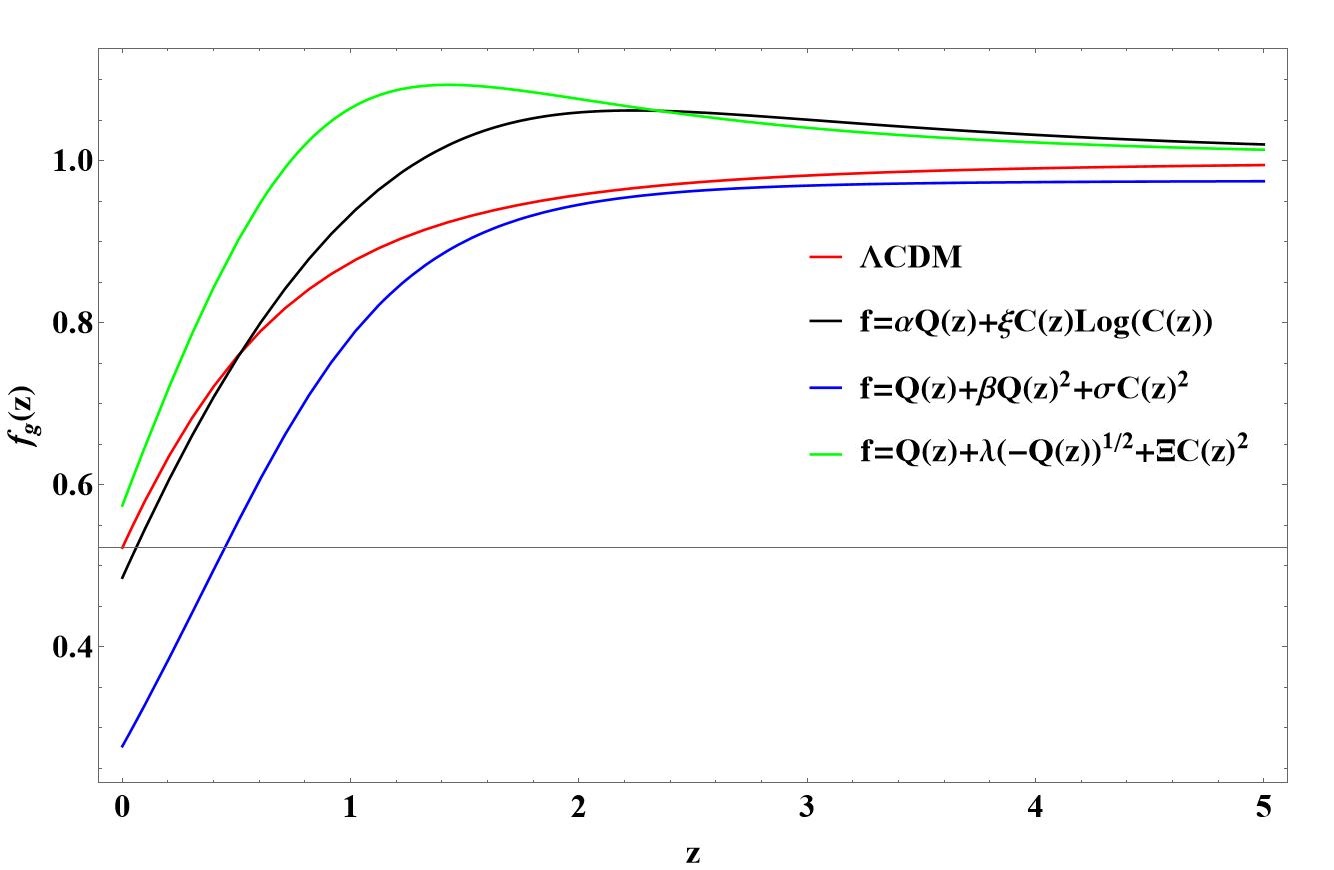}
    \caption{Growth rate $f_g$ vs redshift $z$}
    \label{fig3}
\end{figure}
% From Figure \ref{fig3}, for the models $f(Q, C)=\alpha Q(z)+\xi C(z)\log(c_0C(z))$, $f(Q, C)=Q+\beta Q^2+\sigma C^2$ and $f(Q, C)=Q+\lambda\sqrt{-Q}+\Xi C^2$ the matter growth rate $f_g(z=0)$ are $0.49$, $0.28$ and $0.58$ at present respectively. The matter growth rate in the $\Lambda$CDM $f^{\Lambda CDM}_g(z=0)=0.52$.
}}

The most common parametrization of $f_g$ is in term of $(\Omega^m)^\gamma$ that can be written as
\begin{equation}
    f_g=(\Omega^m)^\gamma\,,\label{grop}
\end{equation}
where $\Omega^m$ and $\gamma$ are functions of scale factor $a$. Here, $\gamma$ is the growth index of matter perturbation. The growth index $\gamma$ is a significant parameter to study the observed large-scale structure (LSS) of the universe to constrain different modified gravity models \cite{nesseris2015, basilakos2016, basilakos2020, kyllep2021, sharma2022, nguyen2023}. By taking the derivative of the logarithm of equation (\ref{grop}), we have 
\begin{equation}
    \frac{d\log f_g}{d\log a}=\left(\log\Omega^m\frac{d\gamma}{d\Omega^m}+\frac{\gamma}{\Omega^m}\right)\frac{d\Omega^m}{d\log a}\,.\label{fgdeq1}
\end{equation}
Similarly, the derivative of equation (\ref{om}) and using equation (\ref{dlnh}), we obtained
\begin{equation}
    \frac{d\Omega^m}{d\log a}=3\omega^{DE}\Omega^m(1-\Omega^m)\,.\label{om1}
\end{equation}
By using equations (\ref{grop}), (\ref{fgdeq1}) and (\ref{om1}), equation (\ref{eveq2}) can be written as
\begin{align}
    3\omega^{DE}\Omega^m(1-\Omega^m)\log\Omega^m\frac{d\gamma}{d\Omega^m}+3\omega^{DE}(1-\Omega^m)&\left(\gamma-\frac{1}{2}\right)+(\Omega^m)^\gamma\notag\\&-\frac{3}{2}\left(\frac{G_{eff}}{G}\right)(\Omega^m)^{1-\gamma}+\frac{1}{2}=0\,.\label{ev1}
\end{align}
The form of equation (\ref{ev1}) is similar to that of other modified gravity theories, such as, $f(\mathring{R})$ gravity discussed in \cite{qing2014}, the main ingredients are encoded in the terms $\omega^{DE}$ and $G_{eff}$. The equation (\ref{ev1}) also yields the growth index parameter $\gamma$, which on contact with observational dataset can give important information on the viability of the $f(Q, C)$ theory. 

\begin{figure}[h!]
     \centering
     \includegraphics[width=\linewidth]{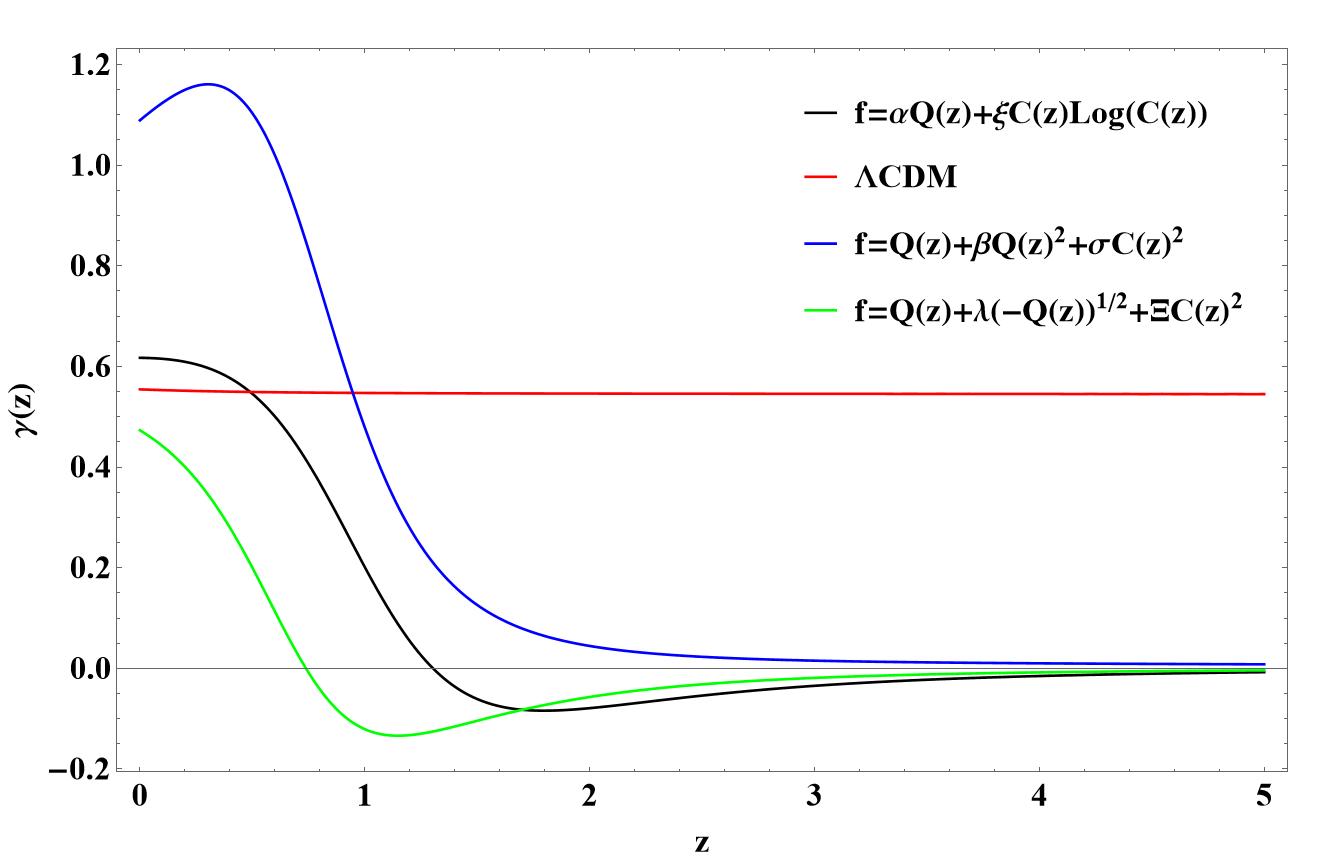}
     \caption{Evolution of growth index $\gamma$ vs redshift $z$}
     \label{fig4}
 \end{figure}

Alternatively, equation (\ref{grop}) can be expressed as \cite{batista2014}
\begin{equation}
\gamma(z)=\frac{\log(f_g)}{\log(\Omega^m)}\,.
\end{equation}
 %and hence using equation (\ref{om})  and the data from Figure \ref{fig3}, 
 We plot evolution of growth index $\gamma(z)$ vs redshift $z$ in Figure \ref{fig4}. Compared with density growth $f_g$ in Figure \ref{fig3}, the growth index $\gamma=\frac{\log {f_g}}{\log{\Omega_m}}>0$ when $f_g,\Omega^m<1$, tends to negative region when $f_g>1$ and $\Omega^m<1$ and tends toward zero when $f_g\approx 1$ and $\Omega^m\approx 0$ at high $z$.
%The growth index $\gamma(z=0)=0.55$ for $\Lambda$CDM in the matter era indicated in Figure \ref{fig4} with the red curve. The growth index $\gamma$ for the models $f=\alpha Q(z)+\xi C(z)\log(c_0C(z))$, $f(Q, C)=Q+\beta Q^2+\sigma C^2$ and $f(Q,C)=Q+\lambda\sqrt{-Q}+\Xi C^2$ are $0.62$, $1.08$ and $0.47$ at the present respectively and  are evolving as shown in the Figure \ref{fig4}.

\section{Conclusion}\label{VII}
\par
In the present work, we have derived the perturbation equations of motion for $f(Q, C)$ gravity, where $Q$ is the non-metricity scalar and $C$ is the boundary term. By calculating the perturbations in this class of modified gravity theories, we have facilitated future investigations on the formation of structures from early to late universe. This study plays a crucial role in determining whether these theories have the potential to challenge the widely accepted $\Lambda$CDM model, particularly in relation to perturbations from both early and late universe perspectives. The resulting field equations (\ref{fefqc11})-(\ref{fefqc44}) are of fourth-order differential equations. The perturbation equation of $f(\mathring{R})$ gravity has been retrieved as demonstrated in the Appendix \ref{fqctofr}. We have used the QS and strong SH approximations to obtain the density contrast evolution equation and the evolution of growth factor $f_g$ for $f(Q, C)$ gravity, the corresponding $\omega^{DE}$ and $G_{eff}$ terms are computed. Finally, by a reparametrisation method, we provide the dynamics of the growth index parameter $\gamma$, which can constrain $f(Q, C)$ gravity models from the CMB and standard siren dataset.

We have considered a few $f(Q,C)$ models e.g., $\alpha Q+\xi C\log(c_0C)$, $Q+\beta Q^2+\sigma C^2$ and $Q+\lambda\sqrt{-Q}+\Xi C^2$ where $c_0=1(\text{km/s/Mpc})^2$, $\alpha=1.3$, $\xi=-2.4$ \cite{kadam2023}, $\beta=0.01 (\text{km/s/Mpc})^2$, $\sigma=-0.1 (\text{km/s/Mpc})^2$, $\lambda=1 (\text{km/s/Mpc})^{-1}$ and $\Xi=1 (\text{km/s/Mpc})^2$ in comparison with $\Lambda$CDM. Figure \ref{fig2}, shows the evolving behaviour of the dark energy equation of state parameter $\omega^{DE}$. %At present, the $\omega^{DE}$ for each of the $f(Q,C)$ models are $-0.65$ and $-0.35$ respectively. Furthermore, from Figure \ref{fig1}, it is observed that the $\delta^m(z=0)=0.85$, $\delta^m(z=0)=0.95$ and $\delta^m(z=0)=0.6$ for the $f(Q,C)$ models considered. 
We have also plotted the density contrast $\delta^m$, the growth rate $f_g$ and the growth index of matter $\gamma$ vs reshift $z$ in Figure \ref{fig1}, Figure \ref{fig3}, Figure \ref{fig4}, respectively. 
The growth index $\gamma(z=0)$ for the first and second models is observed to be higher, while for the third model the growth index is slightly lower than the present value of the growth index $0.55$ of $\Lambda$CDM. This was expected since due to the clustering of dark energy models, the growth index $\gamma$ of linear perturbation can be much lower or higher compared to the $\Lambda$CDM model \cite{batista2014, mehrabi2015a, mehrabi2015, batista2021, avila2022}. When the growth index $\gamma\rightarrow 0$, it means suppression or reversal of matter growth rate. In modified gravity, the enhanced gravitational strength ($G_{eff}>G$) can lead to the growth factor $f_g>1$, which is observed for models i) and iii) for $z> 1.3, 0.8$ respectively from Figure \ref{fig3}, for which the growth index $\gamma<0$ is observed in Figure \ref{fig4}.  Although for $\gamma<0$, from $f_g=(\Omega^m)^\gamma$, $f_g$ increases with decreasing $\Omega^m$. In another way, we can say that the growth rate $f_g$ is slowing down but the $\gamma<0$. This often means the end of growth. Similar behavior is also observed for other modified gravity theories \cite{motohashi2010, linder2019, avila2022,oliveira2024, tsujikawa2007,bellini2014, raveri2014}.
% %%%%%%%%%%%%%%%%%%%%%%%%%%%%%%%%%%%%%%%%%%%%%%%%%%%%%%%%%%%%%%%%%%%%%%%%%%%%%%%%%%%%%%%%%%%%%%%%%%%%%%%%%%%%%%%%%%%%%%%%%%

\appendix \label{appendix1}
\renewcommand{\theequation}{\thesection.\arabic{equation}}
\section{Calculations of perturbation of $f(Q, C)$ gravity}\label{A00}
By considering the vanishing affine connections, $\Gamma^\lambda{}_{\mu\nu}=0$ and $\delta\Gamma^\lambda{}_{\mu\nu}=0$, at first order perturbation we can express

\begin{align}
    &\mathring{\Gamma}^\lambda{}_{\mu\nu}\approxeq\mathring{\bar \Gamma}^\lambda{}_{\mu\nu}+\delta\mathring{\Gamma}^\lambda{}_{\mu\nu}\,, \ \ \ \ \ Q_{\lambda\mu\nu}\approxeq\bar Q_{\lambda\mu\nu} +\delta Q_{\lambda\mu\nu}\,,\ \ \ \ \ Q_\lambda \approxeq\bar Q_\lambda+\delta Q_\lambda\,,\\& \tilde{Q}_\lambda \approxeq\tilde{\bar Q}_\lambda+\delta \tilde{Q}_\lambda,\label{A2} \qquad\qquad P^\lambda{}_{\mu\nu} \approxeq\bar P^\lambda{}_{\mu\nu}+\delta P^\lambda{}_{\mu\nu}\,,\ \ \ \ \ Q \approxeq\bar Q+\delta Q\,,\ \ \ \ \ C \approxeq\bar C+\delta C\,,
\end{align}
where the perturbation terms are
\begin{align}
    \delta\mathring{\Gamma}^\lambda{}_{\mu\nu}=&\frac{1}{2}\left[\delta g^{\sigma\lambda}(\partial_\mu \bar g_{\sigma\nu}+\partial_\nu \bar g_{\mu\sigma}-\partial_\sigma \bar g_{\mu\nu})+\bar g^{\sigma\lambda}(\partial_\mu \delta g_{\sigma\nu}+\partial_\nu \delta g_{\mu\sigma}-\partial_\sigma \delta g_{\mu\nu})\right]=-\delta L^\lambda{}_{\mu\nu}\,,\\
    \delta Q_{\lambda\mu\nu}=&\delta {\mathring{\Gamma}}_{\mu\nu\lambda}+\delta {\mathring{\Gamma}}_{\nu\mu\lambda}\,,
    \delta Q_\lambda=\bar g^{\mu\nu}\delta Q_{\lambda\mu\nu}+\bar Q_{\lambda\mu\nu}\delta g^{\mu\nu}\,,
    \delta\tilde{Q}_\lambda=\bar g^{\lambda\mu}\delta Q_{\lambda\mu\nu}+\bar Q_{\lambda\mu\nu}\delta g^{\lambda\mu}\,,\\
    \delta P^\lambda{}_{\mu\nu}=&\frac{1}{4}\left[2\delta \Gamma^\lambda{}_{\mu\nu}+\delta g_{\mu\nu}(\bar Q^\lambda-\tilde{\bar Q}^\lambda)+\bar g_{\mu\nu}(\delta Q^\lambda-\delta\tilde{Q}^\lambda)-\frac{1}{2}\delta g^{\lambda\sigma}(\bar g_{\sigma\mu}\bar Q_\nu+\bar g_{\sigma\nu}\bar Q_\mu)\right.\notag\\&\left.-\frac{1}{2}\bar g^{\lambda\sigma}(\bar Q_\nu\delta g_{\sigma\nu}+\bar Q_\mu\delta g_{\sigma\nu}+\bar g_{\sigma\mu}\delta Q_\nu+\bar g_{\sigma\nu}\delta Q_\mu)\right]\,,\\
    \delta Q=&\bar Q_{\lambda\mu\nu}\delta P^{\lambda\mu\nu}+\bar P^{\lambda\mu\nu}\delta Q_{\lambda\mu\nu}\,,\qquad\qquad
    \delta C=\delta \mathring{R}-\delta Q\,.\label{A10}
\end{align}

\begin{align}
    \delta (Q f_Q-f)=&f_Q\delta Q-\delta f_Q-\delta f\,,\\
    \delta(f_{QQ} P^\lambda{}_{\mu\nu} \mathring{\nabla}_\lambda Q)=&\delta f_{QQ}\bar{P}^\lambda{}_{\mu\nu}\mathring{\nabla}_\lambda \bar{Q}+f_{QQ}(\delta P^\lambda{}_{\mu\nu})\mathring{\nabla}_\lambda \bar{Q}\notag\\&+f_{QQ}\bar{P}^\lambda{}_{\mu\nu}\delta(\mathring{\nabla}_\lambda Q)\,,\\
    \delta(P^\lambda{}_{\mu\nu}\nabla_\lambda f_C)=&(\delta P^\lambda{}_{\mu\nu})\nabla_\lambda f_C+\bar{P}^\lambda{}_{\mu\nu}\delta (\nabla_\lambda f_C)\,,\\
    \delta\left[\left(\frac{C}{2}g_{\mu\nu}-\mathring{\nabla}_\mu\mathring{\nabla}_\nu+g_{\mu\nu}\mathring{\nabla}^\alpha\mathring{\nabla}_\alpha\right)f_C\right] =&\delta\left(\frac{C}{2}g_{\mu\nu}f_C\right)-\delta(\mathring{\nabla}_\mu\mathring{\nabla}_\nu f_C)+(\delta g_{\mu\nu})\mathring{\nabla}^\alpha\mathring{\nabla}_\alpha f_C\notag\\&+\bar{g}_{\mu\nu}\delta(\mathring{\nabla}^\alpha\mathring{\nabla}_\alpha f_C)\,,
\end{align}
where
\begin{align}
    \delta f=&f_Q\delta Q+f_C\delta C\,,\ \delta f_Q=f_{QQ}\delta Q+f_{QC}\delta C\,,\notag\\&\ \delta f_{QQ}=f_{QQQ}\delta Q+f_{QQC}\delta C\,,\
    \delta(\mathring{\nabla}_\lambda Q)=\partial_\lambda\delta Q\,.
\end{align}
\begin{align}
   \delta(\nabla_\lambda f_C)=&\delta(f_{CC}\mathring{\nabla}_\lambda C)=(\delta f_{CC})\mathring{\nabla}_\lambda \bar{C}+f_{CC}\delta(\mathring{\nabla}_\lambda C)=(\delta f_{CC})\partial_\lambda \bar{C}+f_{CC}\partial_\lambda\delta C\,,\\
   \delta\left(\frac{C}{2}g_{\mu\nu}f_C\right)=&\left(\frac{\delta C}{2}\right)\bar{g}_{\mu\nu}f_C+\frac{\bar{C}}{2}(\delta g_{\mu\nu})f_C+\frac{\bar{C}}{2}\bar{g}_{\mu\nu}\delta f_C\,,\\
   \mathring{\nabla}_\mu\mathring{\nabla}_\nu f_C=&\mathring{\nabla}_\mu(f_{CC}\mathring{\nabla}_\nu \bar{C}+f_{QC}\mathring{\nabla}_\nu \bar{Q})=f_{QQC}\mathring{\nabla}_\mu \bar{Q}\mathring{\nabla}_\nu \bar{Q}+f_{QCC}(\mathring{\nabla}_\mu \bar{Q}\mathring{\nabla}_\nu \bar{C}\notag\\&\qquad\qquad\qquad\qquad\qquad\qquad+\mathring{\nabla}_\nu \bar{Q}\mathring{\nabla}_\mu \bar{C})+f_{CCC}\mathring{\nabla}_\mu \bar{C}\mathring{\nabla}_\nu \bar{C}\,,\\
   \mathring{\nabla}^\alpha\mathring{\nabla}_\alpha f_C=&\bar{g}^{\beta\alpha}(\mathring{\nabla}_\beta\mathring{\nabla}_\alpha f_C)\,.
\end{align}
The perturbations of last two equations are easily derivable from other provided equations.

\section{Scalar metric perturbation of $f(Q,C)$ gravity and $f(\mathring{R})$ gravity}\label{fqctofr}

In this appendix, we provide the detailed calculations of how the perturbation equations of $f(\mathring R)$ theory can be retrieved from equations (\ref{fefqc11})-(\ref{fefqc44}). 

Since $\mathring{R}=Q+C$, by chain rule, we have 
\begin{align}
    &f_Q=f_R \frac{\partial \mathring{R}}{\partial Q}\equiv f_R\,, \qquad f_{QQ}=\frac{\partial}{\partial \mathring{R}}\left(\frac{\partial f}{\partial \mathring{R}}\frac{\partial \mathring{R}}{\partial Q}\right)\frac{\partial\mathring{R}}{\partial Q}\equiv f_{RR}\,,\label{eq46}\\
    &f_C=f_R \frac{\partial \mathring{R}}{\partial C}\equiv f_R\,, \qquad f_{CC}=\frac{\partial}{\partial \mathring{R}}\left(\frac{\partial f}{\partial \mathring{R}}\frac{\partial \mathring{R}}{\partial C}\right)\frac{\partial\mathring{R}}{\partial C}\equiv f_{RR}\,.\label{eq47}
\end{align}
Therefore, equations (\ref{fefqc11})-(\ref{fefqc44}) yield the perturbation equations of $f(\mathring R)$ theories of gravity as
%Hence, using the  reductions given in equations (\ref{eq46}) and (\ref{eq47}), the equations (\ref{fefqc11})-(\ref{fefqc44}) are reduced to well known $f(R)$ gravity equations
\begin{align}
    \kappa^2\delta\rho=&-2\left[3H(H\phi+\dot{\psi})-\frac{\nabla^2\psi}{a^2}+H\frac{\nabla^2B}{a}-H\nabla^2\dot{B}\right]f_R\notag\\&+\left[3H\dot{\delta R}-\left(3H^2+3\dot{H}+\frac{\nabla^2}{a^2}\right)\delta R\right.\notag\\&\left.-18(4H\dot{H}+\ddot{H})\left(2H\phi+\dot{\psi}-\frac{1}{3}\nabla^2\dot{E}\right)-6(4H\dot{H}+\ddot{H})\frac{\nabla^2B}{a}\right]f_{RR}\notag\\&+18H(4H\dot{H}+\ddot{H})f_{RRR}\delta R\,,\label{fefq1}\\
    \kappa^2(\bar\rho+\bar{p}) u=&2\left(H\phi+\dot{\psi}\right)f_R+\left[\dot{\mathring{R}}\phi+H\delta R-\dot{\delta R}\right]f_{RR}-\dot{\mathring{R}}f_{RRR}\delta R\,,\label{fefq2}\\
    \kappa^2\delta p=&2\left[2\dot{H}\phi+H\dot{\phi}+\ddot{\psi}+3H(H\phi+\dot{\psi})+\frac{1}{3}\frac{\nabla^2(\phi-\psi)}{a^2}+\frac{2}{3}H\frac{\nabla^2B}{a}\right.\notag\\&\left.+\frac{1}{3}\frac{\nabla^2\dot{B}}{a}-H\nabla^2\dot{E}-\frac{1}{3}\nabla^2\ddot{E}\right]f_R\notag\\&+\left[2\ddot{\mathring{R}}\phi+\dot{\mathring{R}}\left(4H\phi+\dot{\phi}+2\dot{\psi}+\frac{2}{3}\frac{\nabla^2B}{a}-\frac{2}{3}\nabla^2\dot{E}\right)\right.\\&\left.+\left(3H^2+\dot{H}+\frac{2}{3}\frac{\nabla^2}{a^2}\right)\delta R-2H\dot{\delta R}-\ddot{\delta R}\right]f_{RR}\notag\\&+\left[2\dot{\mathring{R}}^2\phi-(\ddot{\mathring{R}}+2H\dot{\mathring{R}})\delta R-2\dot{\mathring{R}}\dot{\delta R}\right]f_{RRR}-\dot{\mathring{R}}^2f_{RRRR}\delta R\,,\label{fefq3}\\
    -\kappa^2a^2\Pi=0=&\left[\phi-\psi+2aHB+a\dot{B}-3a^2H\dot{E}-a^2\ddot{E}\right]f_R\notag\\&+\left[a\dot{\mathring{R}}(B-a\dot{E})+\delta R\right]f_{RR}\,,\label{fefq4}
\end{align}
where $\mathring{R}=6(2H^2+\dot{H})$ and
\begin{align}
\delta R= &-2\left[6\phi(2H^2+\dot{H})+3H(4\dot{\psi}+\dot{\phi})+3\ddot{\psi}+\frac{\nabla^2(\phi-2\psi)}{a^2}+3H\frac{\nabla^2B}{a}+\frac{\nabla^2\dot{B}}{a}\right.\notag\\&\left.-4H\nabla^2\dot{E}-\nabla^2\ddot{E}\right]\,.
\end{align}
%%%%%%%%%%%%%%%%%%%%%%%%%%%%%%%%%%%%%%%%%%%%%%%%%%%%%%%%%%%%%%%%%%%%%%%%%%%%%%%%%%%%%%%%%%
\section{Scalar metric perturbation of $f(Q, C)$ gravity and $f(Q)$ gravity}\label{A1}
%%%%%%%%%%%%%%%%%%%%%%%%%%%%%%%%%%%%%%%%%%%%%%%%%%%%%%%%%%%%%%%%%%%%%%%%%%%%%%%%%%%%%%%%%%%%%%%%%%%%
For a scalar metric perturbation, the perturbation equations of $f(Q)$ gravity can be deduced from equations (\ref{fefqc11})-(\ref{fefqc44}) by assuming $f$ is independent of $C$:

\begin{align}
    \kappa^2 \delta\rho=&\left[-6H(\dot{\psi}+H\phi)+2\frac{\nabla^2\psi}{a^2}-2H\frac{\nabla^2B}{a}+2H\nabla^2\dot{E}\right]f_Q\notag\\&+\left[72H^3(\dot{\psi}+H\phi)+6H(H^2+\dot{H})\frac{\nabla^2B}{a}-24H^3\nabla^2\dot{E}\right]f_{QQ}\,,\label{fefq11}\\
    \kappa^2 (\bar\rho+\bar{p}) u=&2\left[H\phi+\dot{\psi}\right]f_Q-6H\dot{H}\left[\phi+3\psi+3aHB-\nabla^2E\right]f_{QQ}\,,\label{fefq22}\\
    \kappa^2 \delta p=&\left[2(3H^2+2\dot{H})\phi+2H(\dot{\phi}+3\dot{\psi})+2\ddot{\psi}+\frac{2}{3}\frac{\nabla^2(\phi-\psi)}{a^2}\right.\notag\\&\left.+\frac{2}{3a}(2H\nabla^2B+\nabla^2\dot{B})-\frac{2}{3}(3H\nabla^2\dot{E}+\nabla^2\ddot{E})\right]f_Q\notag\\&
    +\left[-24H^2(3H^2+5\dot{H})\phi-72H(H^2+\dot{H})\dot{\psi}-24H^2(\ddot{\psi}+H\dot{\phi})\right.\notag\\&\left.+24H(H^2+\dot{H})\nabla^2\dot{E}+8H^2\nabla^2\ddot{E}-2H(2H^2+3\dot{H})\frac{\nabla^2B}{a}-2H^2\frac{\nabla^2\dot{B}}{a}\right]f_{QQ}\notag\\&+48H^3\dot{H}\left[6(H\phi+\dot{\psi})+\frac{\nabla^2B}{2a}-2\nabla^2\dot{E}\right]f_{QQQ}\,,\label{fefq33}\\
    -\kappa^2 a^2 \Pi=0=&\left[\phi-\psi+2aHB+a\dot{B}-3a^2H\dot{E}-a^2\ddot{E}\right]f_Q+12H\dot{H}(a^2\dot{E}-aB)f_{QQ}\,.\label{fefq44}
\end{align}
Note that the equations (\ref{fefq11})-(\ref{fefq44}) obtained can be reduced to that of GR\footnote{Equations (\ref{fefq11})-(\ref{fefq44}) are reduced to equations as in \cite{weinberg2008, sriram2012} when we set $f=Q$.}. The equations obtained in \cite{jimenez2020} cannot be reduced to the same after converting conformal time to cosmic time, especially the $\delta p$ expression. Furthermore, all the $f_{QQ}$ coefficients are different than ours.

% Note that the equations (\ref{fefq11}) and (\ref{fefq33}) can be expressed in much simpler form in term of Bardeen potentials.
%%%%%%%%%%%%%%%%%%%%%%%%%%%%%%%%%%%%%%%%%%%%%%%%%%%%%%%%%%%%%%%%%%%%%%%%%%%%

\end{document}